\documentclass[reprint,amsmath,amssymb,aps,prb,]{revtex4-2}

\usepackage{graphicx}
\usepackage{dcolumn}
\usepackage{bm}
\usepackage{physics}
\usepackage{multirow}
\usepackage{color}
\usepackage{soul}

\begin{document}

\title{Tight-binding model with sublattice-asymmetric spin-orbit coupling\\for square-net nodal line Dirac semimetals}

\author{Gustavo S.~Orozco-Galvan,$^{1}$ Amador Garc\'ia-Fuente$^{2,3}$ and Salvador Barraza-Lopez$^{1}$}
\affiliation{1. Department of Physics, University of Arkansas, Fayetteville, AR 72701, USA and MonArk NSF Quantum Foundry, University of Arkansas, Fayetteville, AR 72701, USA\\
2. Departamento de F\'{i}sica, Universidad de Oviedo, E-33007 Oviedo, Spain\\
3. Centro de Investigaci\'{o}n en Nanomateriales y Nanotecnolog\'{\i}a, Universidad de Oviedo–CSIC, 33940 El Entrego, Spain}
\date{\today}

\begin{abstract}
We study a 4-orbital tight-binding (TB) model for ZrSiS from the square sublattice generated by the Si atoms. { After studying three other alternatives}, we endow such model with a new effective spin-orbit coupling (SOC) consistent with {\em ab initio} dispersions around the Fermi energy ($E_F$) in four systematic steps: (1) We calculate the electronic dispersion of bulk ZrSiS using an implementation of density-functional theory (DFT) based on numeric atomic orbitals [{\em J. Phys.: Condens. Matter} {\bf 14}, 2745 (2002)] in which on-site and off-site SOC can be told apart. As a result, we determine that local SOC-induced band gaps around $E_F$ are predominantly created by the on-site contribution. (2) Gradually reducing the atomic basis set size, we then create an electronic band structure with 16 orbitals per unit cell (u.c.) which retains the qualitative features of the dispersion around $E_F$, including SOC-induced band gaps. (3) Zr is the heaviest element on this compound and it has a non-negligible contribution to the electronic dispersion around $E_F$; we show that it provides the strongest contribution to the SOC-induced band gap. (4) Using L\"{o}wdin partitioning approach, we project the effect of SOC onto the 4-orbital Hamiltonian. This way, we facilitate an effective SOC interaction that was explicitly informed by {\em ab initio} input.
\end{abstract}

\maketitle

\section{\label{sec:Intro}Introduction}

Topological semimetals (TSMs) are a class of materials whose crossings between valence and conduction bands are topologically protected \cite{gao2019topological}. There are two main subcategories of TSMs: nodal point semimetals (NPSs), and nodal line semimetals (NLSs). NPSs have a discrete set of crossing points at $E_F$, while NLSs  display a continuous crossing line. Initial, theory-lead searches for NLSs called for materials with nonsymmorphic symmetry \cite{young2015dirac,kruthoff2017topological} such as ZrSiS \cite{schoop2016dirac}, ZrSiSe, and ZrSiTe \cite{Jin}. Those materials display unusually-high magnetoresistance \cite{singha2017large,sankar2017crystal,ali2016butterfly,voerman2019origin,wang2016evidence}, high carrier mobilities \cite{zhang2018transport,sankar2017crystal,schilling2017flat}, and linear band dispersions around $E_F$~\cite{schoop2016dirac,singha2017large,fu2019dirac}.

\begin{figure}[tb]
\centering
\includegraphics[width=0.4\textwidth]{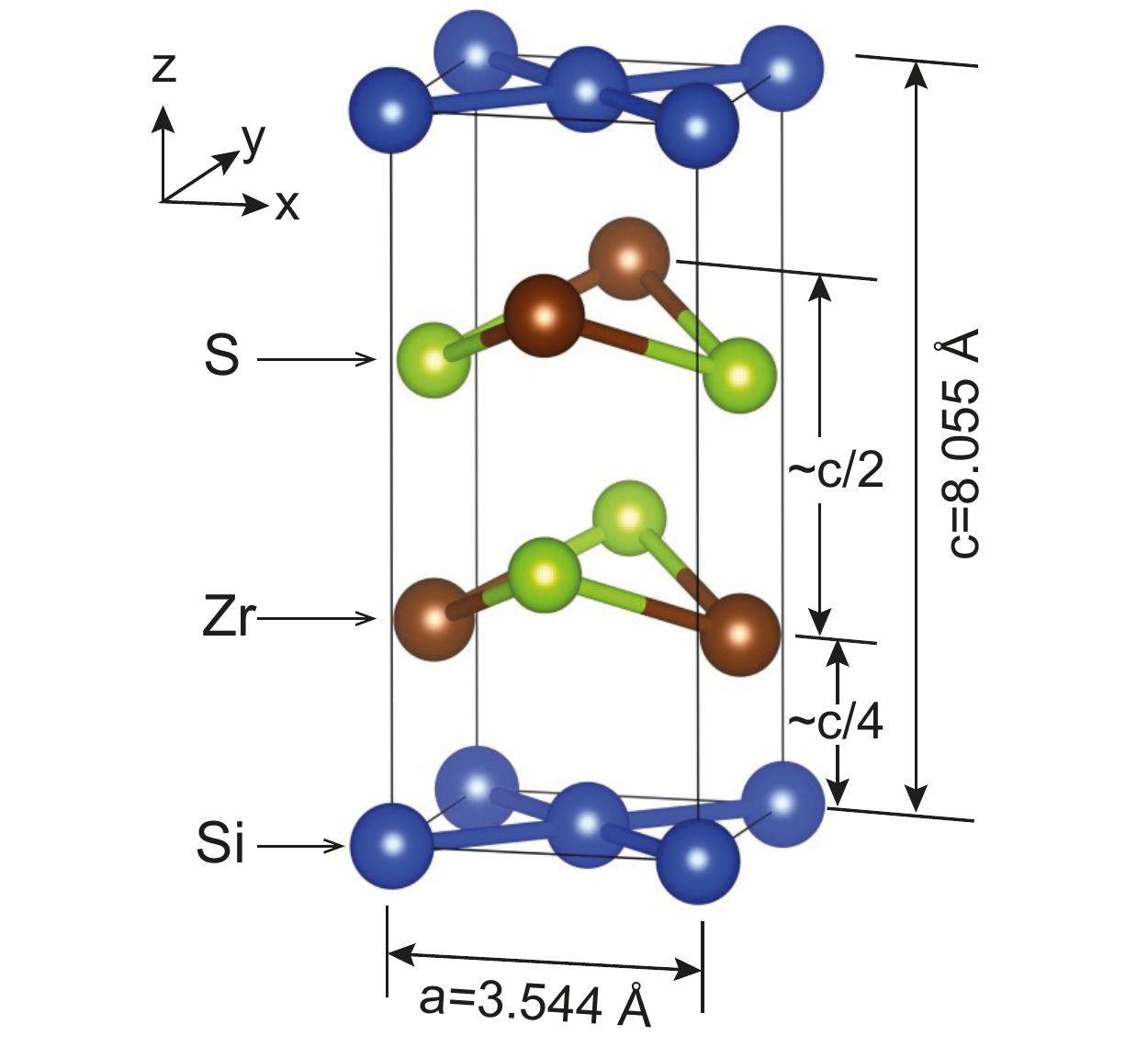}
\caption{\label{fig:f1}Tetragonal u.c.~for ZrSiS. Zr, Si and S atoms are represented by brown, blue and green spheres, respectively. Si square nets are seen at the bottom and top of the u.c. Lattice parameters are also shown.}
\end{figure}

\begin{figure}[tb]
\centering
\includegraphics[width=0.48\textwidth]{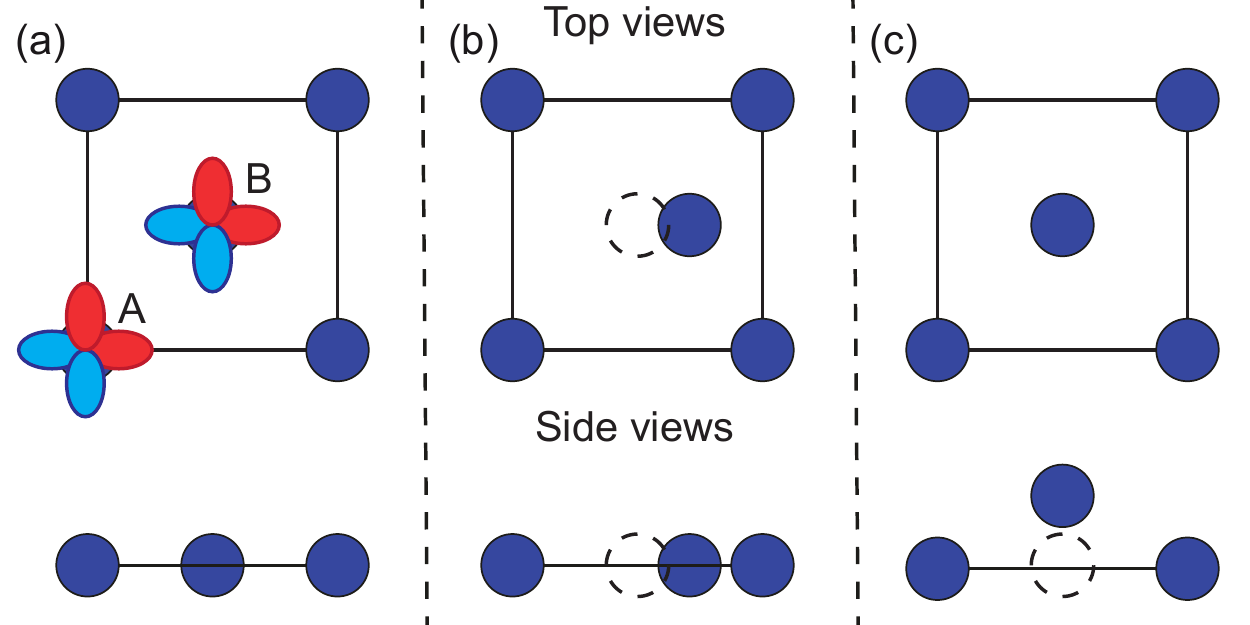}
\caption{\label{fig:pxpymodels} U.c.s for the $p_xp_y$ models presented in Refs.~\cite{klemenz2020systematic,luo2015room} and \cite{deng2022twisted}: (a) The u.c.~in Klemenz {\em et al.} is formed by two equivalent sites and two orbitals ($\ket{p_x}$ and $\ket{p_y}$, explicitly drawn) per site. (b) The model proposed for Deng and coworkers relies on square nets whose B atom is horizontally displaced. (c) Buckled square lattice, where the B atom is vertically displaced, as proposed by Luo and Xiang. (The SOC in Aryal model was introduced without an atomistic justification on a symmetric lattice like the one shown in (a), and no model is drawn here for that reason.)}
\end{figure}

ZrSiX (X=S, Se, or Te) belongs to the nonsymmorphic P4/nmm space group. As shown in Fig.~\ref{fig:f1}, two atoms of each species are present in their u.c., with their Si atoms arranged in stacked square nets~\cite{teicher20223d,banerjee2021higher,michen2022mesoscopic,rosmus2022electronic,he2021impurity,herrera2023tunable,cano2021moire}.  The electronic structure of two-site square lattices is described by $\ket{p_x}$ and $\ket{p_y}$ orbitals in various models. Indeed, such a basis set was proposed by Luo and Xiang to study a Bi-based topological insulator \cite{luo2015room}. Later, Klemenz and coworkers discovered nodal line semimetals in square nets~\cite{klemenz2019topological,klemenz2020role,klemenz2020systematic}, which are 2D {\em symmorphic} lattices, and presented a symmetry analysis of such nets relying on  $\ket{p_x}$ and $\ket{p_y}$ orbitals in Ref.~\cite{klemenz2020systematic} (Fig.~\ref{fig:pxpymodels}(a)). Since then, other two additional teams have implemented two-site {\em $p_xp_y$} models to study the effect of SOC in similar nets~\cite{deng2022twisted,aryal2022topological}.

Recent quantum oscillation experiments have shown that the Berry phase generated for ZrSiS depends on the orientation of the magnetic field \cite{ali2016butterfly,voerman2019origin} and Yang {\em et al.}~suggested that a key agent in such dependence is SOC~\cite{yang2021anisotropic}. It is natural to include SOC in $p_xp_y$ models to understand the observations made in Ref.~\cite{yang2021anisotropic}. It is not possible, nevertheless, to gap the crossings at the Fermi level in these models upon the customary implementation of SOC~\cite{konschuh2010tight}. Luo and Xiang \cite{luo2015room}, Deng {\em et al.}~\cite{deng2022twisted}, and Aryal {\em et al.}~\cite{aryal2022topological} independently realized that $p_xp_y$ square net models can be gapped at zero energy through a combination of (i) an {\em on-site SOC} (this is, one in which neighboring atoms do not contribute to SOC), and (ii) a {\em lowering of symmetry such that the two atoms in the u.c.~turn inequivalent}. Such asymmetry was created by a relative displacement of one atom along the 2D plane (Fig~\ref{fig:pxpymodels}(b))~\cite{deng2022twisted}, or by an out-of-plane relative displacement (buckling, Fig~\ref{fig:pxpymodels}(c)) \cite{luo2015room}. On the other hand, Aryal {\em et al.} incorporated an on-site asymmetry between sublattices directly onto the SOC interaction {\em ad hoc} (meaning that no microscopic justification was offered to motivate their choice for SOC interaction).

{ Given that the Si square nets on ZrSiS,  ZrSiSe, and ZrSiTe are in fact {\em not distorted} in the way done in Refs.~\cite{luo2015room} and \cite{deng2022twisted}, and that Aryal's work did not provide justification for their SOC}, {\em we seek the source of an asymmetric SOC that could be added to the $p_xp_y$ models to produce a band gap at zero energy}, and thus make it consistent with {\em ab initio} results containing SOC. The search for that mechanism from {\em ab initio} information is our guiding principle, and the motivation for this work.

We have previously worked with 2D models for topological insulators relying on $\ket{p_x}$ and $\ket{p_y}$ orbitals also~\cite{barraza2022two}, and we possess a deep knowledge of {\em ab initio-}informed TB formulations~\cite{martin_2020} as provided in the SIESTA code \cite{soler2002siesta,Junquera}. We have written extensions to this code for quantum transport \cite{PhysRevLett.104.076807} and {\em ab initio} molecular dynamics  \cite{PhysRevLett.117.246802} calculations. We are also closely familiar with SOC on this code, {\em down to implementing it} \cite{PhysRevLett.102.246801,fernandez2006site,cuadrado2021validity,caracedo}. Our knowledge of TB models \cite{PhysRevB.87.155436,Naumis_2017,Barraza-Lopez_2023} makes the extraction of TB parameters \cite{slater1954simplified} from SIESTA possible, to the point of turning on-site SOC interactions for analysis at will. This experience underpins the description of a SOC-induced mechanism to break sublattice symmetry.

\begin{figure*}[tb]
\centering
\includegraphics[width=0.96\textwidth]{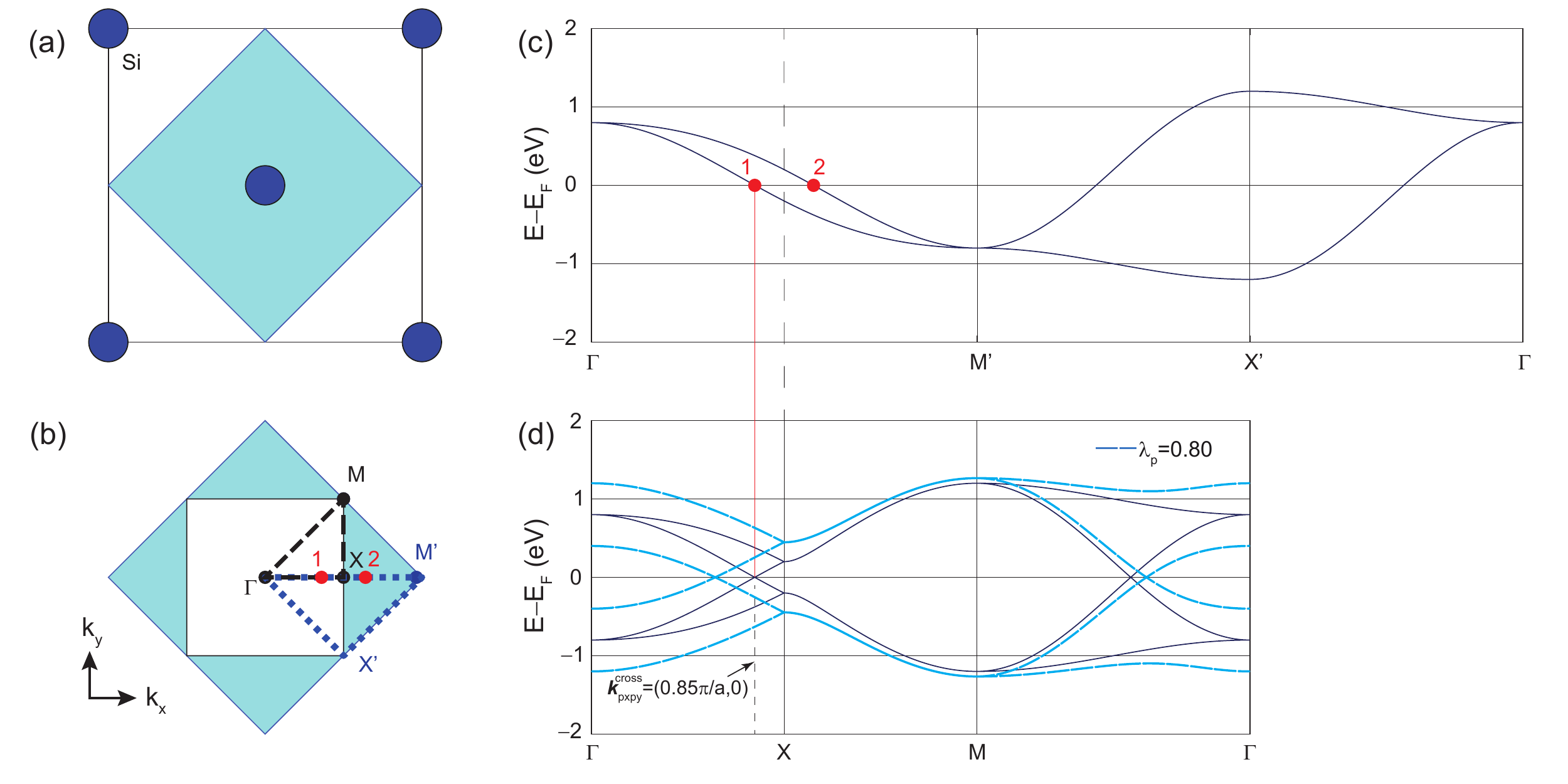}
\caption{\label{fig:f2}Band structure of the $p_xp_y$ model \cite{klemenz2020systematic}. (a) Single-atom u.c.~(blue) and original u.c.~containing two Si atoms (shown within black lines). (b) First Brillouin zone for the u.c.~containing a single Si atom (blue, rotated square) and folded first Brillouin zone (white square) corresponding to the u.c.~containing two Si atoms.
The $M'$ point folds into $\Gamma$, while the $X'$ and $M$ points are equivalent. The $X$ point sits midway from the $\Gamma$ and $M'$ points. Points $1$ and $2$ sit at equal distances away from the middle of the $\Gamma-M'$ segment. (c) Electronic dispersion generated by the unfolded Hamiltonian; Eqn.~\eqref{eq:H2}. (d) Dispersion generated by the Hamiltonian on the folded first Brillouin zone; Eqn.~\eqref{eq:H4} { (the dashed cyan curve is obtained with the SOC due to p-orbitals, Eqn.~\eqref{eq:socreference})}. The red line linking subplots (c) and (d) indicates that bands crossing point $2$ in subplot (c) fold into bands crossing point $1$ in subplot (d). The dashed line passing through (c) corresponds to the $X$ point in the folded Brillouin zone; subplot (d). The horizontal axes on subplots (c) and (d) were chosen for a direct comparison of crystal momentum: this is, we kept proportional lengths for unfolded and folded band structures.}
\end{figure*}

The article is organized as follows: Sec.~\ref{sec:KSC} contains a brief { overview of the  $p_xp_y$ models, and three previous implementations of SOC on them}. We demonstrate that the leading contribution of SOC on ZrSiS is due to on-site interactions in Sec.~\ref{sec:Num}. To simplify the model, an on-site SOC will be used from that point onwards. A DFT-based TB model is developed in Sec~\ref{sec:TB} by a gradual removal of numerical atomic orbitals from the basis set; we determine the orbitals that contribute to the electronic dispersion of the material around $E_F$ the most, while preserving the SOC band gaps. We also determine Zr to contribute to the on-site SOC the most. Applying the L\"{o}wdin partitioning scheme to this auxiliary TB model, we effectively induce SOC-gaps into the $p_xp_y$ model provided in Ref.~\cite{klemenz2020systematic} that break sublattice symmetry in Sec.~\ref{sec:Low}. Conclusions are presented in Sec.~\ref{sec:Con}.

\section{\label{sec:KSC}$p_xp_y$ models}

\subsection{Band folding in $p_xp_y$ models}

The u.c.~of $p_xp_y$ models is represented by the 2D square net depicted in Figs.~\ref{fig:pxpymodels}(a) and \ref{fig:f2}(a), formed by one atom at the corner (A) and another at the center (B). The equivalence among A and B sublattices permits an analysis of the models within the single-atom u.c.~(shaded blue square in Fig.~\ref{fig:f2}(a)). Fig.~\ref{fig:f2}(b) depicts the first Brillouin zones associated to the two u.c.s. The first Brillouin zone of the two-atom u.c.~(white square in Fig.~\ref{fig:f2}(b)) is related to the first Brilluoin zone of the single atom u.c.~(shaded blue square) by folding. Here, primed $k-$points ($k_x',k_y'$) refer to the unfolded Brillouin zone, while unprimed $k-$points ($k_x,k_y$) describe high symmetry points on the folded Brillouin zone. An exception is the $\Gamma$ point, which is the same for both, and left unprimed for that reason.

Comparison of the unfolded and folded first Brillouin zones in Fig.~\ref{fig:f2}(b) shows the equivalence among the $X'$ and $M$ points, the folding of the $M'$ point onto the $\Gamma$ point, and the fact that the $X$ point of the folded Brillouin zone sits halfway among the $\Gamma$ and $M'$ points of the unfolded one. We next analyze the consequences of previous observations on the electronic dispersion within the $p_xp_y$ models.

The electronic dispersion within the single-atom u.c.~is given by the $2\times 2$ matrix:
\begin{align}
\label{eq:H2}
&H_{pxpy}^{2\times 2}(k_x',k_y') =
\nonumber\\
&\left(
\begin{smallmatrix}
2t_\sigma^{(1)}\cos({k_x'a'})+2t_\pi^{(1)}\cos({k_y'a'}) & 2(t_\sigma^{(2)}-t_\pi^{(2)})\sin({k_x'a'})\sin({k_y'a'}) \\
2(t_\sigma^{(2)}-t_\pi^{(2)})\sin({k_x'a'})\sin({k_y'a'}) & 2t_\sigma^{(1)}\cos({k_y'a'})+t_\pi^{(1)}\cos({k_x'a'})
\end{smallmatrix}
\right),
\end{align}
where $a'=a/\sqrt{2}$.
This Hamiltonian matrix is written in the basis of rotated orbitals ($\ket{p_x'}=\frac{1}{\sqrt{2}}(\ket{p_x}+\ket{p_y})$, and $\ket{p_y'}=\frac{1}{\sqrt{2}}(\ket{p_x}-\ket{p_y})$). Table~\ref{ta:pxpymod} contains the values of the TB parameters $t_\sigma^{(1)}$, $t_\pi^{(1)}$, $t_\sigma^{(2)}$ and $t_\pi^{(1)}$, adapted from Refs.~\cite{luo2015room,klemenz2020systematic,deng2022twisted,aryal2022topological}. This electronic dispersion contains two bands (or four when spin degeneracy is considered) and it is shown in Fig.~\ref{fig:f2}(c), where the parameters from Ref.~\cite{klemenz2020systematic} have been used.

\begin{table}
\centering
\caption{ TB parameters (in eV) for the $p_xp_y$ models, adapted from Refs.~\cite{luo2015room,klemenz2020systematic,deng2022twisted}, and \cite{aryal2022topological}. $\lambda_p$ is the intensity of SOC in units of eV, $\epsilon$ is an effective parameter in units of eV arising from the asymmetry of the lattice, and $\delta<1$ is the dimensionless distortion used in Ref.~\cite{deng2022twisted}.\label{ta:pxpymod}}
\begin{tabular}{c|cccc|ccc}
\hline
\hline
Model & $t_\sigma^{(1)}$ & $t_\pi^{(1)}$ & $t_\sigma^{(2)}$ & $t_\pi^{(2)}$ & $\lambda$ & $\epsilon$ & $\delta$\\
\hline
Klemenz {\em et al.} \cite{klemenz2020systematic} &  0.50 &  $-$0.10 &  0.05   & $-$0.05 & --    & --   & --\\
Aryal 	{\em et al.} \cite{aryal2022topological}  &  1.90 &  $-$0.50 & $-$0.10 & 0.10    & 0.20  & 0.10 & --\\
Luo 	{\em et al.} \cite{luo2015room}           &  1.00 &  $-$0.20 &  0.00   & 0.00    & 1.20  & 0.40 & -\\
Deng 	{\em et al.} \cite{deng2022twisted}       &$-$2.00&     0.70 & $-$0.40 & 0.14    & 0.02  & --   & 0.10\\
\hline
\hline
\end{tabular}
\end{table}

The band structure in the 2-atom u.c.~is constructed on a basis of four orbitals: $\{\ket{p_x^A}$, $\ket{p_y^A}$, $\ket{p_x^B}$, $\ket{p_y^B}\}$, where $A$ and $B$ label the two sites within the u.c. In this basis, the general form of the $4\times 4$ matrix Hamiltonian is
\begin{align}
\label{eq:H4}
H_{pxpy}^{4\times 4}(k_x,k_y) =
\begin{pmatrix}
\alpha_x & 0 & \beta & \gamma \\
0 & \alpha_y & \gamma  & \beta  \\
\beta & \gamma & \alpha_x & 0\\
\gamma & \beta & 0 & \alpha_y
\end{pmatrix},
\end{align}
where:
\begin{align}
\label{eq:H4x4}
\alpha_x &\equiv 2t_\sigma^{(2)}\cos({k_xa}) + 2t_\pi^{(2)}\cos({k_ya}),
\nonumber\\
\alpha_y &\equiv 2t_\sigma^{(2)}\cos({k_ya}) + 2t_\pi^{(2)}\cos({k_xa}),
\nonumber\\
\beta &\equiv (t_\sigma^{(1)}+t_\pi^{(1)})\left(\cos(\frac{k_x+k_y}{2}a) + \cos(\frac{k_x-k_y}{2}a)\right),\nonumber\\&\text{ and}
\nonumber\\
\gamma &\equiv (t_\sigma^{(1)}-t_\pi^{(1)})\left(\cos(\frac{k_x+k_y}{2}a) - \cos(\frac{k_x-k_y}{2}a)\right).
\end{align}


Increasing the size of the u.c.~decreases the size of its first Brillouin zone by the same proportion (Fig.~\ref{fig:f2}(b)). This implies that, for every band within the original 2D Brillouin zone, there are two bands in the folded one. Fig.~\ref{fig:f2}(c) and Fig.~\ref{fig:f2}(d) show the electronic dispersions of $H_{pxpy}^{2\times 2}(k_x',k_y')$ and $H_{pxpy}^{4\times 4}(k_x,k_y)$, following the $k$-point trajectories marked with blue and black in Fig.~\ref{fig:f2}(b), respectively. The crucial point is that the linear crossing at zero energy taking place at $\mathbf{k}_{pxpy}^{cross}=(0.85\pi/a$,0) along the $\Gamma-X$ line in Fig.~\ref{fig:f2}(d) is due to band folding, as a consequence of the equivalence among the $A$ and $B$ sublattices.

{

\subsection{SOC and asymmetric sublattices}

On-site SOC in $p_xp_y$ models can be expressed in terms of Pauli matrices $\tau$, $\upsilon$, and $\sigma$ acting respectively on the orbital $\{\ket{p_x},\ket{p_y}\}$ site $\{A,B\}$, and spin $\{+,-\}$ spaces:
\begin{align}\label{eq:socreference}
H_{SOC} = \frac{\lambda_p}{2}\tau_y\otimes\upsilon_0\otimes\sigma_z,
\end{align}
where $\upsilon_0$ is the identity matrix in the sublattice space. This SOC interaction does not open an energy gap at zero energy because it does not induce an inequivalence among sites, {\em i.~e.} still allows for folding (see dashed curves in Fig.~\ref{fig:f2}(d) and Supplemental Material~\footnote{Supplemental Material contains a description of the on-site SOC, a MATLAB program to reproduce the bands of the 16-orbital model and to obtain the parameter $\eta$ from {L\"{o}wdin} partitioning technique, and a TB electronic dispersion of slabs using the 16-orbital model.}).

To overcome this shortcoming, Aryal {\em et al.}~proposed the explicit introduction of an on-site asymmetry, $H_{Aryal}$, that together with $H_{SOC}$ would open a gap~\cite{aryal2022topological}:
\begin{align}\label{eq:Aryal}
H_{Aryal} = \epsilon\tau_x\otimes\upsilon_z\otimes\sigma_z,
\end{align}
where $\epsilon$ is a constant with units of energy. Since $H_{Aryal}$ is proportional to $\upsilon_z$, the two atoms in the u.c.~become inequivalent and folding is now forbidden, thus, a gap is open under on-site SOC.

Similarly, Luo and Xiang introduced a perturbation that incorporates effective hopping terms that arise from the buckling of the lattice~\citep{luo2015room}:
\begin{align}\label{eq:LX}
H_{L\&X} &= -\epsilon\sin\left(\frac{k_x-k_y}{2}a\right)\tau_y\otimes\upsilon_y\otimes\sigma_x +
\nonumber\\
&+\epsilon\sin\left(\frac{k_x+k_y}{2}a\right)\tau_y\otimes\upsilon_y\otimes\sigma_y.
\end{align}

Lastly, the Hamiltonian proposed by Deng and coworkers can be expressed as the Hamiltonian of the undistorted lattice $H_{pxpy}^{4\times4}\otimes\sigma_0$ (with $\sigma_0$ being the identity matrix in spin space) plus an asymmetric sublattice perturbation. Such perturbation acquires the form:
\begin{align}\label{eq:Deng}
H_{Deng} &= \operatorname{Re}\left((h_1-\beta)\tau_0 + h_2\tau_z + (h_3-\gamma)\tau_x\right)\otimes\upsilon_x\otimes\sigma_0 +
\nonumber\\
&-\operatorname{Im}\left((h_1-\beta)\tau_0 + h_2\tau_z + (h_3-\gamma)\tau_x\right)\otimes\upsilon_y\otimes\sigma_0,
\end{align}
where
\begin{align}
h_1 &= \left(t_1e^{i\frac{k_xa}{2}} + t_2e^{-i\frac{k_xa}{2}}\right)\cos\left(\frac{k_ya}{2}\right)e^{-i\delta k_xa},
\nonumber\\
h_2 &= \left(t_3e^{i\frac{k_xa}{2}} + t_4e^{-i\frac{k_xa}{2}}\right)\cos\left(\frac{k_ya}{2}\right)e^{-i\delta k_xa},
\nonumber\\
h_3 &= \left(2it_5e^{i\frac{k_xa}{2}} - 2it_6e^{-i\frac{k_xa}{2}}\right)\sin\left(\frac{k_ya}{2}\right)e^{-i\delta k_xa},
\end{align}
and
\begin{align}
t_1 &= 2\left(\frac{a}{2d_1}\right)^4\left((1-2\delta)^2+1\right)(t_\sigma^{(1)}+t_\pi^{(1)}),
\nonumber\\
t_2 &= 2\left(\frac{a}{2d_2}\right)^4\left((1+2\delta)^2+1\right)(t_\sigma^{(1)}+t_\pi^{(1)}),
\nonumber\\
t_3 &= 2\left(\frac{a}{2d_1}\right)^4\left((1-2\delta)^2-1\right)(t_\sigma^{(1)}-t_\pi^{(1)}),
\nonumber\\
t_4 &= 2\left(\frac{a}{2d_2}\right)^4\left((1+2\delta)^2-1\right)(t_\sigma^{(1)}-t_\pi^{(1)}),
\nonumber\\
t_5 &= 2\left(\frac{a}{2d_1}\right)^4(1-2\delta)(t_\sigma^{(1)}-t_\pi^{(1)})
\nonumber\\
t_6 &= 2\left(\frac{a}{2d_2}\right)^4(1+2\delta)(t_\sigma^{(1)}-t_\pi^{(1)})
\nonumber\\
d_1 &= \sqrt{\left(\frac{a}{2}\right)^2 + \left(\frac{a}{2} - \delta a\right)^2}
\nonumber\\
&\text{ and}
\nonumber\\
d_2 &= \sqrt{\left(\frac{a}{2}\right)^2 + \left(\frac{a}{2} + \delta a\right)^2}.
\end{align}

In contrast to $H_{Aryal}$, where a sublattice dependent interaction was added on-site, the distortion of the lattice (either in-plane or out of plane) produces first nearest neighbor hoppings which are odd under sublattice symmetry, giving rise to terms proportional to $\upsilon_y$. Even though those SOC perturbations are different, all of them are characterized by {\em breaking sublattice symmetry}. We note in passing that we determined their SOC couplings to apply to the unprimed basis introduced here (this is why $\tau_x$ was used in \eqref{eq:Aryal} instead of $\tau_z$ as reported in Ref.~\cite{aryal2022topological}), and that the explicit expressions for $H_{L\&X}$ and $H_{Deng}$ as written in Eqns.~\eqref{eq:LX} and \eqref{eq:Deng} did not exist in the original sources but were written for an unified discussion and a straight comparison among models and our results which, as seen in next Sections, will turn out to be different.

}

%

\section{\label{sec:Num}Predominant on-site SOC on ZrSiS NLS}

\begin{figure*}[t]
\centering
\includegraphics[width=0.96\textwidth]{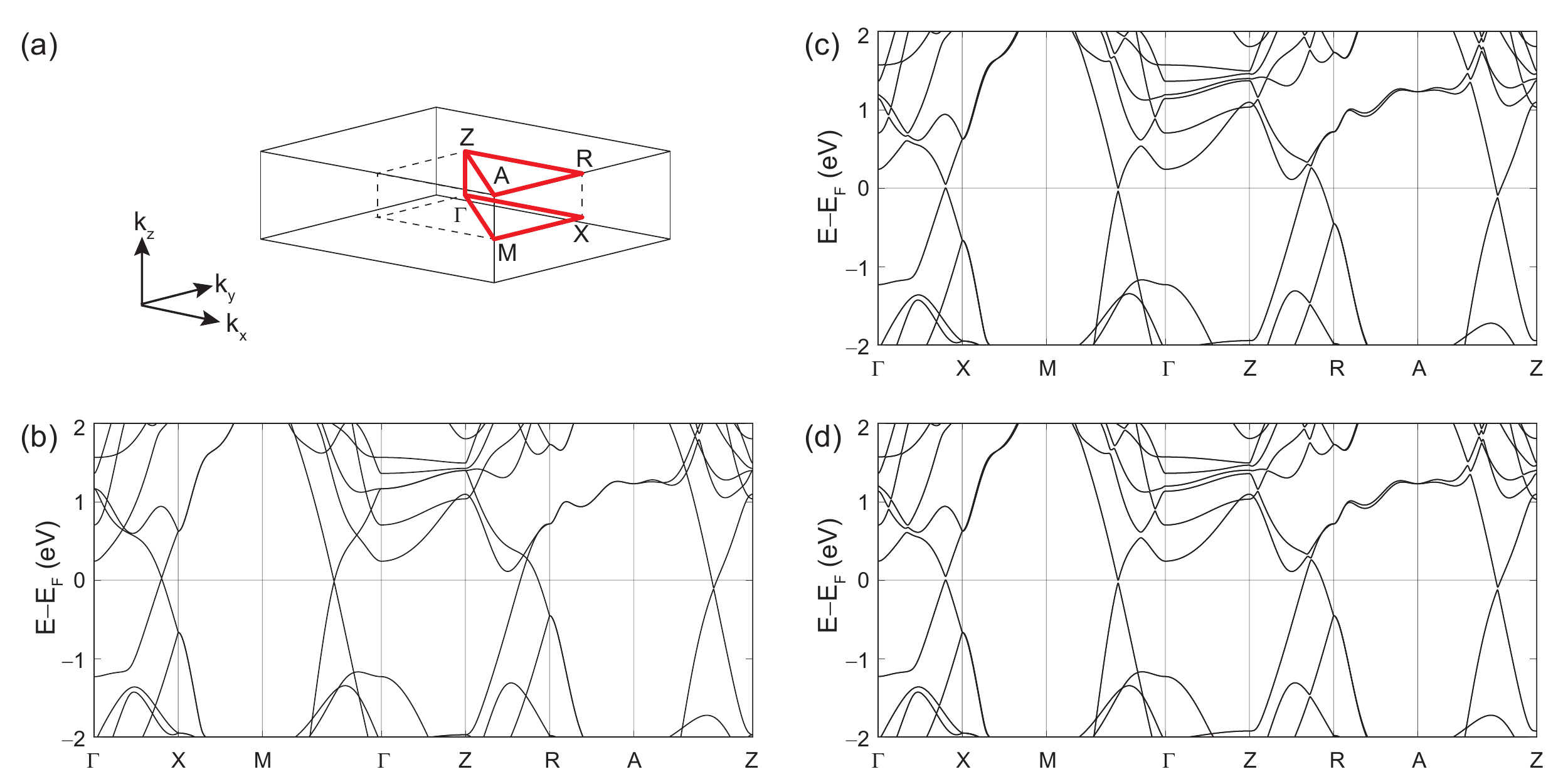}
\caption{\label{fig:f_3n}(a) First Brillouin zone for ZrSiS. (b) Electronic band structure for ZrSiS without SOC. (c) Electronic structure with on-site SOC. (d) Electronic structure  under full SOC. All band structures on this figure were obtained using a standard (DZP) basis set containing 82 (164) orbitals in the unit cell without (with) SOC.}
\end{figure*}

Although SOC is not constrained to be an on-site effect~\cite{fernandez2006site,kurita2020systematic}, Cuadrado \textit{et al.}~have shown that this approximation is usually sufficient \cite{cuadrado2021validity}. We shall carry a detailed study of SOC in ZrSiS relying on the DFT SIESTA code \cite{soler2002siesta}. The code produces TB representations of the electronic structure with Hamiltonian matrix elements calculated at the {\em ab initio} level onto a basis of numeric atomic-like orbitals~\cite{Junquera}. This TB representation is generated without the need for post-processing (read Wannierization).

A so-called double-$\zeta$-plus polarization basis set (DZP, or standard) is used in SIESTA calculations \cite{Junquera}. It contains two radial functions per angular momentum in the valence, and one extra (``polarizing'') channel for an unoccupied (excited) orbital momentum channel. For Si and S, the DZP basis set includes two orbitals for the $s-$channel, six for the $p-$channel, and five for the (polarizing) $d-$channel, leading to thirteen atomic-like orbitals for each of those two species \cite{Junquera}. For the Zr atom, the basis contains two radial functions for the $s-$ and $d-$channels, and one radial function for the $p-$channel, leading to 15 orbitals per atom. This way, ZrSiS is described with a basis set containing 82 localized orbitals.

SIESTA uses relativistic pseudopotentials \cite{cuadrado2012fully} and SOC can be introduced in two gradual ways: either (i) through the on-site approximation, which only couples orbitals within the same atom \cite{fernandez2006site}, or (ii) by including matrix elements that couple orbitals belonging to the same atom but also to other atoms, {\em i.e.} off-site terms. We dubbed this second approach a {\em full} SOC.

The electronic dispersion of ZrSiS with and without SOC corrections and a standard (DZP) basis set is depicted in Fig.~\ref{fig:f_3n}. The first Brillouin zone is displayed in Fig.~\ref{fig:f_3n}(a), Fig.~\ref{fig:f_3n}(b) contains an electronic dispersion without SOC, while Fig.~\ref{fig:f_3n}(c) and Fig.~\ref{fig:f_3n}(d) are band structures with on-site and full SOC, respectively. Visual comparison of Figs.~\ref{fig:f_3n}(c) and \ref{fig:f_3n}(d) indicates that the predominant contribution of SOC comes from on-site interactions (in particular, the gaps at the $\Gamma-X$ segment for the on-site and full SOC are both approximately $44$~meV).

Fig.~\ref{fig:f_3n} shows that contributions beyond the on-site interaction are small already, but we shall be more rigorous and {\em obtain a band structure associated to a system only perturbed by the off-site SOC} next. Even though SIESTA does not explicitly provide this option, it stores Hamiltonian and overlap matrices which can thus be examined and manipulated at will. We obtain the off-site SOC dispersion by defining its associated Hamiltonian and overlap matrices as:
\begin{align}
H^{offsite} &= H^{full}-H^{onsite}+H_0\otimes\sigma_0,
\nonumber\\
S^{offsite} &= S^{full}-S^{onsite}+S_0\otimes\sigma_0,
\end{align}
where $H_0$ ($S_0$), $H^{onsite}$ ($S^{onsite}$), $H^{offsite}$ ($S^{offsite}$) and $H^{full}$ ($S^{full}$) are the Hamiltonian (overlap) matrices without SOC, with on-site SOC, under off-site, and under full SOC, respectively.

\begin{figure}[t]
\centering
\includegraphics[width=0.48\textwidth]{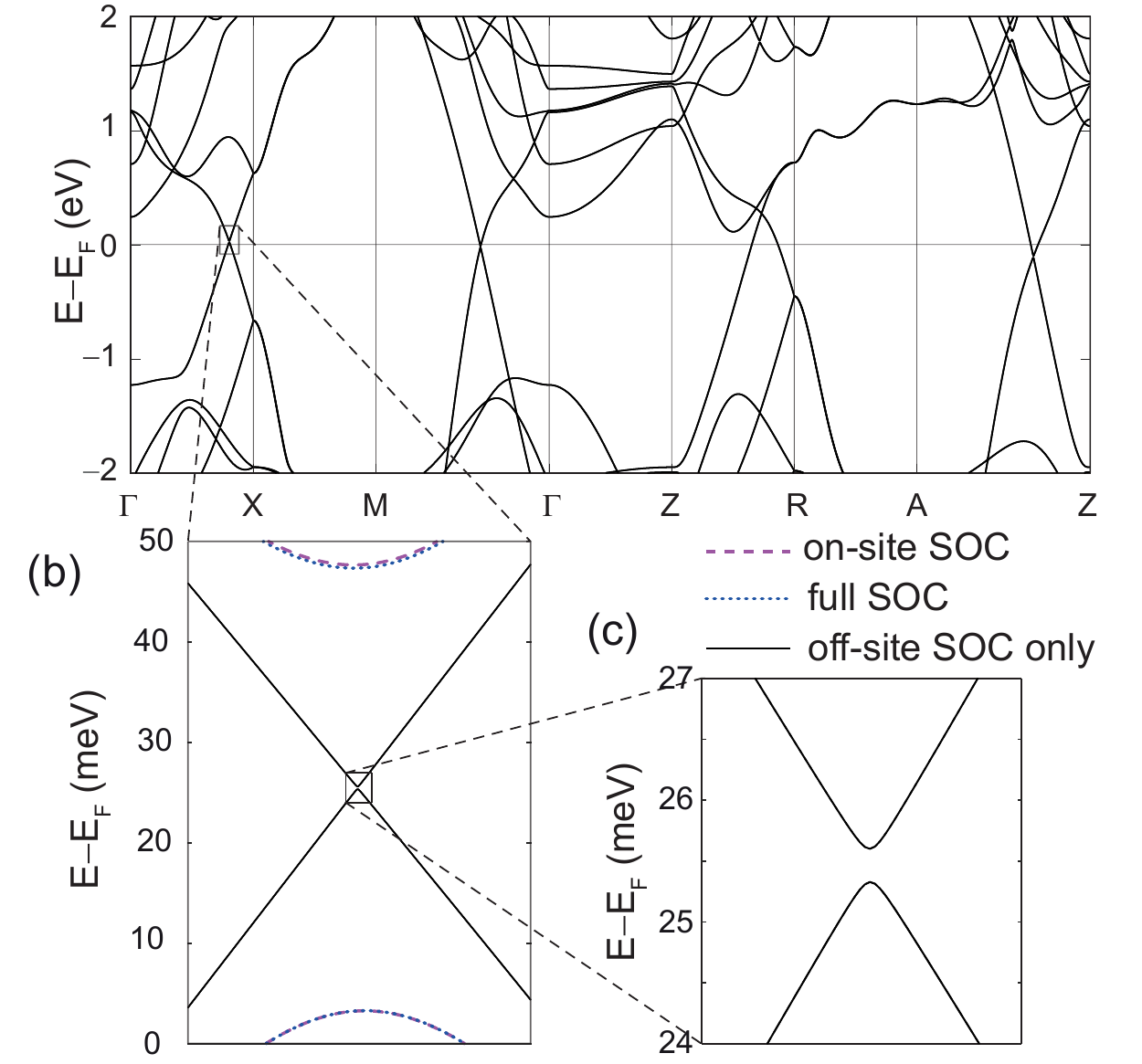}
\caption{\label{fig:f_4n}(a) Band structure of ZrSiS under off-site SOC only. (b) Zoom-in of the vicinity of the crossing along $\Gamma-X$. The dispersions associated to on-site SOC and full SOC were included in purple and blue for comparison. (c) Second zoom-in: the off-site SOC-gap is smaller than $0.3$ meV.}
\end{figure}

Fig.~\ref{fig:f_4n}(a) depicts the band structure when only the off-site SOC is turned on. Fig.~\ref{fig:f_4n}(b) is a zoom-in around the vicinity of the crossing along the $\Gamma-X$ segment, where the bands with on-site and full SOC have been included for comparison. The energy gap with just off-site SOC is $\sim0.3$~meV, a value {\em two orders of magnitude smaller than the gap of roughly 20 meV reported in Refs.~\cite{schoop2016dirac,hosen2017tunability} and the one seen in Fig.~\ref{fig:f_3n}}. Consequently, the gap is due to the on-site SOC predominantly.

The fact that the band gaps around $E_F$ in ZrSiS NLS are predominantly due to an on-site SOC invites to extend the basis of the $p_xp_y$ model beyond Si $\ket{p_x}$ and $\ket{p_y}$ orbitals, which will permit observing SOC-gaps without lowering the symmetry of the Si square net~\cite{luo2015room,deng2022twisted}, and without the need to postulate a SOC interaction that breaks sublattice symmetry {\em ad-hoc} \cite{aryal2022topological}. We will develop a TB Hamiltonian with less than 82 orbitals (but more than the four ones used in the $p_xp_y$ models) in Sec.~\ref{sec:TB}. That TB model will reproduce SOC-induced energy gaps due to an on-site SOC. We will embed this SOC into the $p_xp_y$ model through L\"{o}wdin partitioning method subsequently (Sec.~\ref{sec:Low}).

\section{\label{sec:TB}DFT-based TB models with basis sets smaller than DZP to isolate effective on-site SOC effects}

\subsection{Rationale}
We now design a DFT-based TB description for ZrSiS to identify the orbitals involved in the SOC-induced band gap openings. As indicated in Sec.~\ref{sec:Num}, the on-site SOC will suffice, and it will be utilized from now on.

\begin{figure*}[t]
\centering
\includegraphics[width=0.96\textwidth]{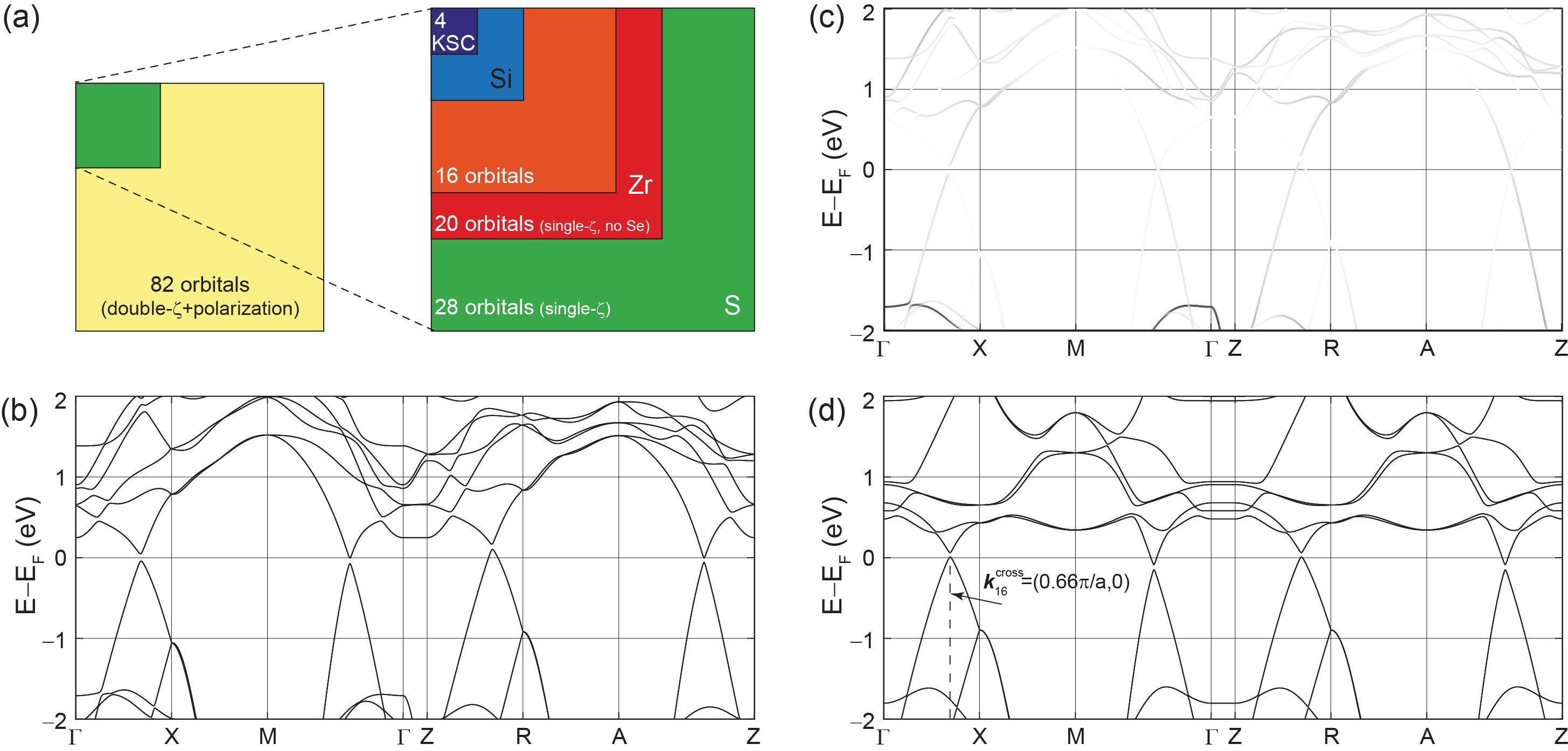}
\caption{\label{fig:f_5n}(a) Hamiltonian size {\em versus} the basis type (the size doubles upon inclusion of SOC). (b) Band structure without SOC turned on, using a (28-orbital) SZ basis set. (c) Band structure projected onto S atoms. Note the minimal hybridization around $E_F$, which justifies removing those from the description. (d) Band structure from the 16-orbital set with on-site SOC turned on; the location of the avoided band crossing, $\mathbf{k}_{16}^{cross}=(0.66\pi/a,0)$, was indicated.}
\end{figure*}

One route toward TB models from DFT involves {\em post-processing} of a Hamiltonian originally written on a plane-wave basis set, into a basis of maximally localized Wannier functions \cite{Wannier,RevModPhys.84.1419}. We follow another (similar) approach here: starting with the DFT-Hamiltonian written in a localized basis consisting of atomic orbitals \cite{soler2002siesta,Junquera}, we reduce the number of orbitals and their radial extent gradually. This will shed light into SOC-induced sublattice symmetry breaking on NLSs in this material family, which is the point of this manuscript.

\subsection{Reducing the size of the orbital basis set}

As displayed in Table \ref{ta:basissets}, our initial band structure calculation for ZrSiS relied on 82 atomic orbitals, and the Hamiltonian doubled in size when SOC was included. We wish to determine orbitals with leading contributions to the on-site SOC and--as seen in Table \ref{ta:basissets}--our search included a single$-\zeta$ (SZ) basis set for Zr, Si, and S (which contains one radial function per valence channel, and no polarizing orbitals) with 28 orbitals, a SZ set for Zr and Si only (which effectively removed S from the electronic structure) with 20 orbitals, and a smaller orbital set with 16 orbitals, in which the $\ket{s}$ and $\ket{d_{z^2}}$ orbitals of Zr were removed. The basis set of $p_xp_y$ models containing four orbitals is shown in Table \ref{ta:basissets} as well. Fig.~\ref{fig:f_5n}(a) represents the gradual size reduction of the (square) Hamiltonian matrices achieved as the basis sets were reduced in size.

The band structure shown in Fig.~\ref{fig:f_5n}(b) was created with the (28-orbital) SZ basis set. The radial cutoffs of the numerical atomic orbitals \cite{Junquera} were set to 2.12 \AA{} to admit second nearest-neighbor interactions at most. The overlap matrix was approximated as the identity matrix in order to reduce the number of parameters on the TB representation.  The on-site SOC was included on this Hamiltonian [41], leading to the gapped electronic dispersion around $E_F$. The crucial observation is that the inclusion of additional orbitals to the $p_xp_y$ Hamiltonian set leads onto gap openings by on-site SOC only.

\begin{table}
\centering
\caption{Size of Hamiltonian {\em versus} basis set employed. The listed size takes into account the existence of two atoms per atomic species in the u.c. The DZP (standard) basis set was used to obtain the band structures in Fig.~\ref{fig:f_3n}, which serve as the starting point (reference) for our study. The $p_xp_y$ Hamiltonian listed here led to Fig.~\ref{fig:f2}, and it corresponds to the folded scenario (containing two Si atoms in the u.c.).\label{ta:basissets}}
\begin{tabular}{c|ccc|c}
\hline
\hline
Basis set name & Zr & Si & Se & Size of Hamiltonian \\
\hline
DZP            & 15 & 13 & 13 & 82\\
SZ             &  6 &  4 &  4 & 28\\
20-orbital set &  6 &  4 &  0 & 20\\
16-orbital set &  4 &  4 &  0 & 16\\
$p_xp_y$      &  0 &  2 &  0 &  4\\
\hline
\hline
\end{tabular}
\end{table}

The TB Hamiltonian can be further simplified, while still preserving its crucial qualitative features. The SZ basis set for ZrSiS contains 14 orbitals (Si $\to$ $\ket{s}$, $\ket{p_x}$, $\ket{p_y}$, $\ket{p_z}$; Zr $\to$ $\ket{s}$, $\ket{d_{xy}}$, $\ket{d_{yz}}$, $\ket{d_{z^2}}$, $\ket{d_{xz}}$, $\ket{d_{x^2-y^2}}$; S $\to$ $\ket{s}$, $\ket{p_x}$, $\ket{p_y}$, $\ket{p_z}$), yielding 28 orbitals per u.c. To discard orbitals, we computed their contribution to the energy bands around $E_F$. This was done by determining the eigenvectors for every band at each $k$-point. Fig.~\ref{fig:f_5n}(c) displays the projection of S orbitals onto the electronic dispersion of the 28 orbital model with on-site SOC. The darker the hue, the stronger the contribution. S orbitals barely contribute to the states within Fig.~\ref{fig:f_5n}(c) and hence can be removed from the basis set without a significant distortion of the low energy electronic dispersion. Analogously, further removal of the $\ket{s}$ and $\ket{d_{z^2}}$ orbitals of Zr still leads to a gapped band structure around $E_F$ when the on-site SOC is turned on; Fig.~\ref{fig:f_5n}(d) depicts the band structure of the resulting (16-orbital) TB electronic dispersion. The avoided band crossing takes place at $\mathbf{k}_{16}^{cross}=(0.66\pi/a,0)$.

\begin{figure}[t]
\centering
\includegraphics[width=0.48\textwidth]{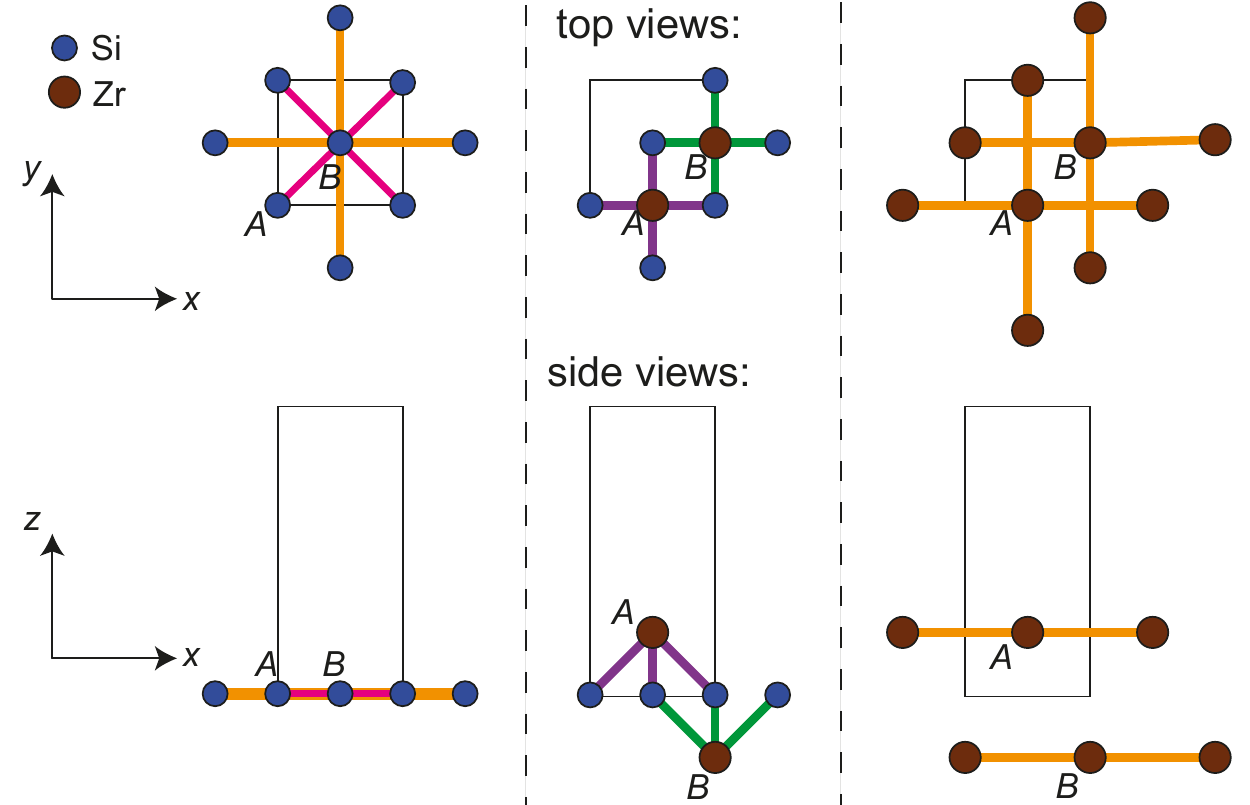}
\caption{\label{fig:AC}Top and side views of atomic connections. From left to right, Si-Si, Zr-Si and Zr-Zr connections are displayed. Orange bonds indicate intralayer hopping of orbitals separated by an in-plane lattice constant, while pink represents hoppings among atoms closer than one in-plane lattice constant. Green and purple lines correspond to interlayer hopping between ZrB and Si, and ZrA and Si atoms, respectively.}
\end{figure}

{
This 16-orbital model provides a qualitative understanding of the origin of SOC in ZrSiS: Fig.~\ref{fig:AC} illustrates our labelling convention. One Si $A$ atom is located at the origin of coordinates, at the corner of the prism in Fig.~\ref{fig:f1}. The Si $B$ atom lies at the center of the base. Each Si layer is ``sandwiched'' by two layers of Zr symmetrically separated. The top Zr layer (above the Si layer) is made of $A$ atoms, while Zr $B$ atoms conform the bottom layer (below the Si layer). Colored lines in Fig.~\ref{fig:AC} link the atoms whose interaction is taken into account for this model (Appendix~\ref{appendixA}). Orange and pink lines link atoms located at the same layer, purple links Zr's $A$ atom with the Si layer below it, while green links Zr's $B$ atom with the Si layer above it. As we can observe, the same chemical species only interact with each other when they are located at the same layer: Zr's $A$ and $B$ atoms do not interact with each other. Interlayer hopping only happens between Zr atoms and the closest layer of Si atoms.

Furthermore, for a given pair of Si atoms separated by an in-plane lattice constant ({\em i.e.} belonging to the same sublattice), there is an associated Zr atom whose in-plane projection lies halfway between the two Si. When the Si atoms belong to the A sublattice, and their relative separation is along the x-direction ($\pm a\hat{x}$), their associated Zr is above the Si layer. In contrast, if such pair of Si atoms belongs to the B sublattice, the associated Zr would be below. {\em It is this difference what produces the sublattice asymmetry} discussed in Sec.~\ref{sec:KSC} for the case of ZrSiS.}

\begin{figure}[tb]
\centering
\includegraphics[width=0.48\textwidth]{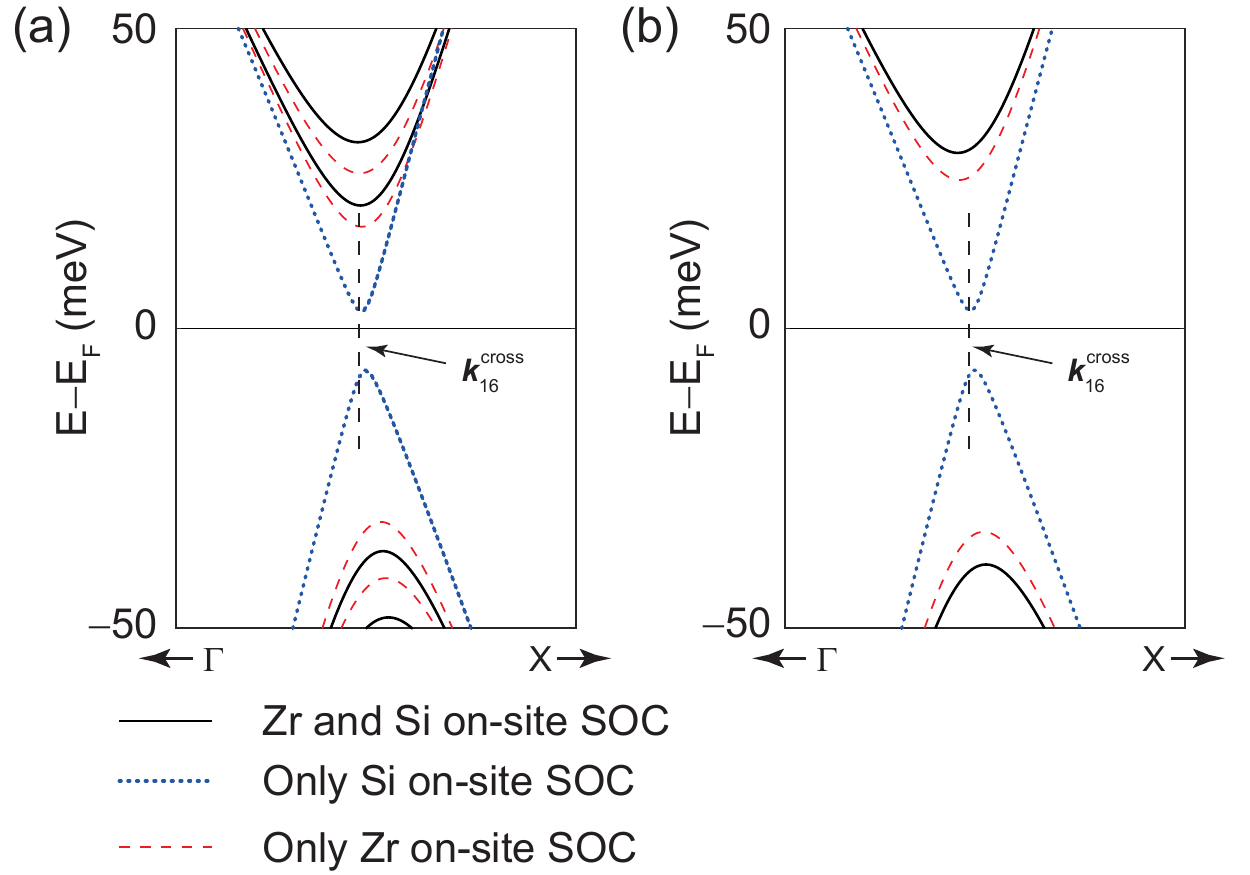}
\caption{\label{fig:f_6} (a) Zoom-in to band-gap at $E_F$ induced by on-site SOC in the $\Gamma-X$ segment. Black, blue, and red bands include the contribution of both (Zr and Si) on-site SOC, only Si on-site SOC, and only Zr on-site SOC. Complete on-site SOC was considered ({\em i.e.} matrix elements coupling orbitals with the same spin and with different spins). (b) Dispersion when only orbitals with opposite spins contribute remain degenerate, as in the DFT results of Figs.~\ref{fig:f_3n} and \ref{fig:f_5n}.}
\end{figure}

\subsection{Spin components}

At this point, we have identified the main atomic orbitals involved in the gap opening process, however, we can still be more specific and determine the atom whose on-site SOC contributes the most. Fig.~\ref{fig:f_6}(a) displays a zoom-in around the band gap generated by on-site SOC (black bands). The red and blue bands are obtained when only Zr or Si on-site SOC contributions are considered, respectively. Since Si on-site SOC gap is $\sim6$ times smaller than the gap seen in Fig.~\ref{fig:f_3n}(d), we can neglect its contribution to simplify the model. {  This is an important difference among the model that we are developing and the models discussed in Sec.~\ref{sec:KSC}: {\em The SOC produced by the atoms in the square net is not required to gap the Fermi level}. Rather, the on-site Zr SOC suffices}.

Furthermore, on-site SOC couples orbitals with the same spin ({\em i.e.}~through matrix elements of the form $\bra{\phi,\pm}H_{SOC}\ket{\varphi,\pm}$) or with opposite spins ($\bra{\phi,\pm}H_{SOC}\ket{\varphi,\mp}$) [41]. Fig~\ref{fig:f_6}(b) is a zoom-in at the on-site gap generated by recourse to only matrix elements that couple orbitals with opposite spins. Analogously to Fig~\ref{fig:f_6}(a), in Fig~\ref{fig:f_6}(b) red, blue, and black bands arise from Zr on-site SOC, Si on-site SOC, and both (Zr and Si) on-site SOC, respectively. The coupling among opposite spins is responsible for the band splitting at $E_F$.

The SOC due to the four Zr $d-$orbitals in the 16-band model looks as follows:
\begin{equation}\label{eq:ZrSOC}
H^{Zr}_{SOC}=\frac{\lambda_d}{2}
\left(
\begin{smallmatrix}
0 & 0 & 0 & 2i & 0 & -1  & -i & 0 \\
0 & 0 & i & 0 & 1 & 0 & 0 & -i \\
0 & -i & 0 & 0 & i & 0 & 0 & 1 \\
-2i & 0 & 0 & 0 & 0 & i & -1 & 0 \\
0 & 1 & -i & 0 & 0 & 0 & 0 & -2i \\
-1 & 0 & 0 & -i & 0 & 0 & -i & 0 \\
i & 0 & 0 & -1 & 0 & i & 0 & 0 \\
0 & i & 1 & 0 & 2i & 0 & 0 & 0 \\
\end{smallmatrix}
\right).
\end{equation}

The matrix entries in Eqn.~\eqref{eq:ZrSOC} were ordered as follows: $\{\ket{d_{xy},+}$, $\ket{d_{yz},+}$, $\ket{d_{xz},+}$, $\ket{d_{x^2-y^2},+}$, $\{\ket{d_{xy},-}$, $\ket{d_{yz},-}$, $\ket{d_{xz},-}$, and $\ket{d_{x^2-y^2},-}\}$. The important point is that, unlike $\ket{p_x}$ and $\ket{p_y}$ orbitals in Eqn.~\eqref{eq:socreference}, {\em on-site SOC in Zr does mix spin components}. As shown in Fig.~\ref{fig:f_6}, {\em such coupling of opposite spins is behind the band gap opening seen in DFT calculations}.

The 16-orbital Hamiltonian (Si: $\ket{s}$, $\ket{p_x}$, $\ket{p_y}$, $\ket{p_z}$; Zr: $\ket{d_{xy}}$, $\ket{d_{yz}}$, $\ket{d_{xz}}$, $\ket{d_{x^2-y^2}}$) strikes a balance among basis size and a proper qualitative description of the electronic dispersion with on-site SOC. The parameters for this TB Hamiltonian are provided in Appendix A, and our code is provided in the Supplemental Material. We will consider only Zr on-site SOC among orbitals with opposite spins in next section.

\section{\label{sec:Low}Adding SOC into the $p_xp_y$ model from a L\"odwin partitioning of the 16-orbital Hamiltonian}

L\"{o}wdin partitioning will now be used \cite{lowdin1963studies,roland2003spin} to project the on-site SOC due to Zr within the 16-orbital model onto the (4-orbital) Si $p_xp_y$ Hamiltonian at the vicinities of their crossing points ($\mathbf{k}_{16}^{cross}$ and $\mathbf{k}_{p_xp_y}^{cross}$, respectively). The first steps are a diagonalization without SOC of both the 16-orbital Hamiltonian ($H_{16}\otimes\sigma_0$) and the $p_xp_y$ one ($H_{pxpy}^{4\times 4}\otimes\sigma_0$), and a rearrangement of the diagonalized Hamiltonians, such that the two double degenerate bands around $E_F$ (that we call ``low-energy'' bands) appear at the four uppermost entries:
\begin{align}
&\mathcal{H}_{p_xp_y}(\mathbf{k}_{p_xp_y}^{cross}+\Delta\mathbf{k})=\nonumber\\
&\begin{pmatrix}
\mathcal{H}_{p_xp_y}^{low}(\mathbf{k}_{p_xp_y}^{cross}+\Delta\mathbf{k}) & O \\
O & \mathcal{H}_{p_xp_y}^{high}(\mathbf{k}_{p_xp_y}^{cross}+\Delta\mathbf{k}) \\
\end{pmatrix}, \text{ and}
\nonumber\\
&\mathcal{H}_{16}(\mathbf{k}_{16}^{cross}+\Delta\mathbf{k}) =\nonumber\\
&\begin{pmatrix}
\mathcal{H}_{16}^{low}(\mathbf{k}_{16}^{cross}+\Delta\mathbf{k}) & O \\
O & \mathcal{H}_{16}^{high}(\mathbf{k}_{16}^{cross}+\Delta\mathbf{k}) \\
\end{pmatrix},
\end{align}
(the curly $\mathcal{H}$ emphasizes that those Hamiltonians without SOC are all diagonal, and $\Delta\mathbf{k}$ is a small excursion away from the $k-$points where the crossing takes place). The point is for the ``low-energy blocks", $\mathcal{H}_{p_xp_y}^{low}$ and $\mathcal{H}_{16}^{low}$, to describe the same two double degenerate bands despite them belonging to different models:
\begin{align}
\mathcal{H}_{p_xp_y}^{low}(\mathbf{k}_{p_xp_y}^{cross}+\Delta\mathbf{k}) \approx \mathcal{H}_{16}^{low}(\mathbf{k}_{16}^{cross}+\Delta\mathbf{k}),
\end{align}
up to a scaling factor.

\begin{figure}[t]
\centering
\includegraphics[width=0.48\textwidth]{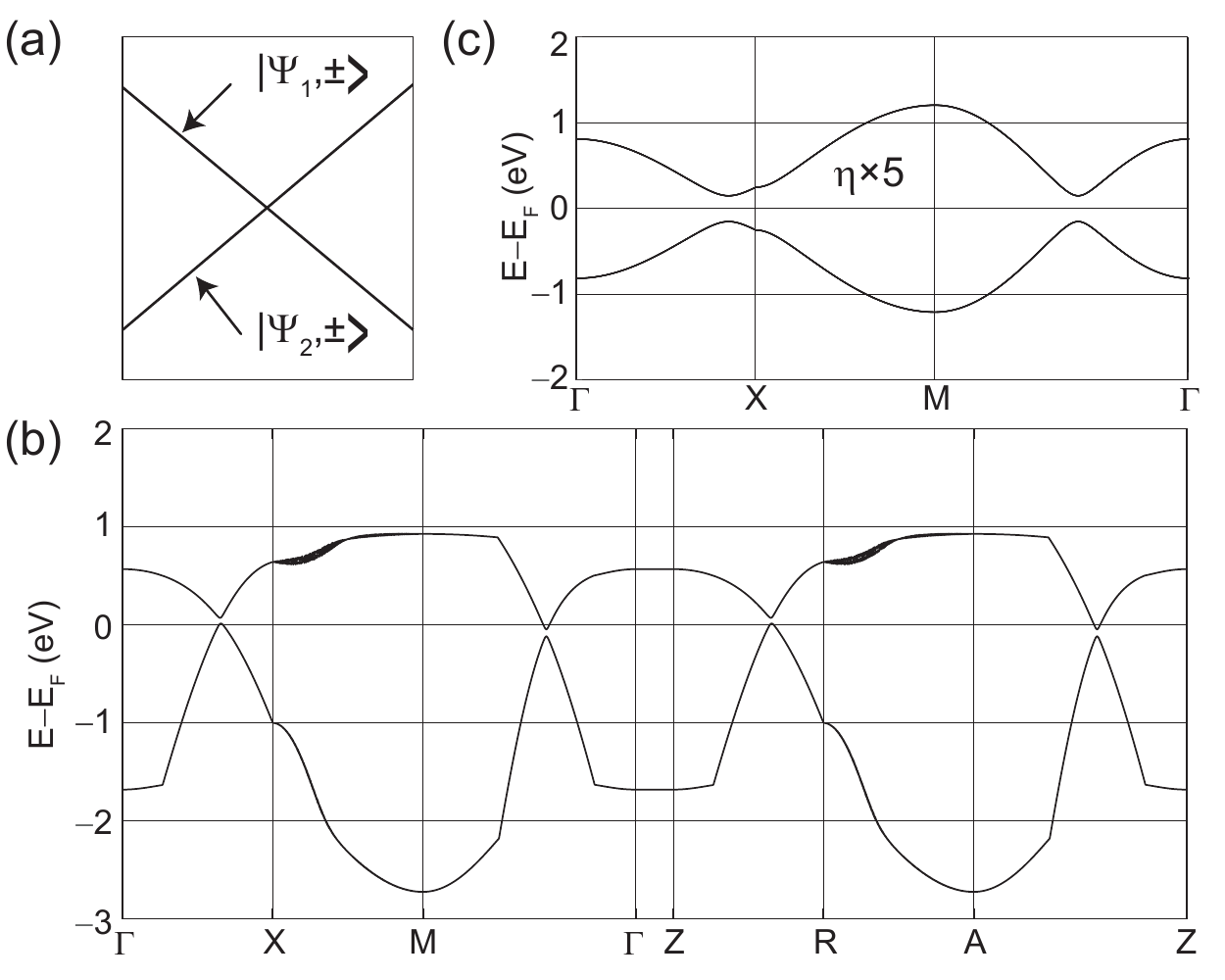}
\caption{\label{fig:f_7n}(a) Schematic definition of degenerate eigenvectors $\ket{\psi_1,\pm}$ and $\ket{\psi_2,\pm}$ associated to low-energy bands when SOC is off. (b) Electronic dispersion resulting from the perturbed low-energy 16-orbital Hamiltonian with on-site Zr SOC: $\mathcal{H}_{16}^{low}+\mathcal{P}^{low}$. (c) Electronic dispersion of the $p_xp_y$ Hamiltonian under the induced SOC: $\mathcal{H}_{p_xp_y}^{low}+\mathcal{P}^{low}$; we increased the value of $\eta$ by a factor of five to better resolve bands in Fig.~\ref{fig:f_8n}. (There is band splitting at the $X-M$ segment near 1 eV that does not affect our analysis nor conclusions.)}
\end{figure}

\begin{figure}[t]
\centering
\includegraphics[width=0.48\textwidth]{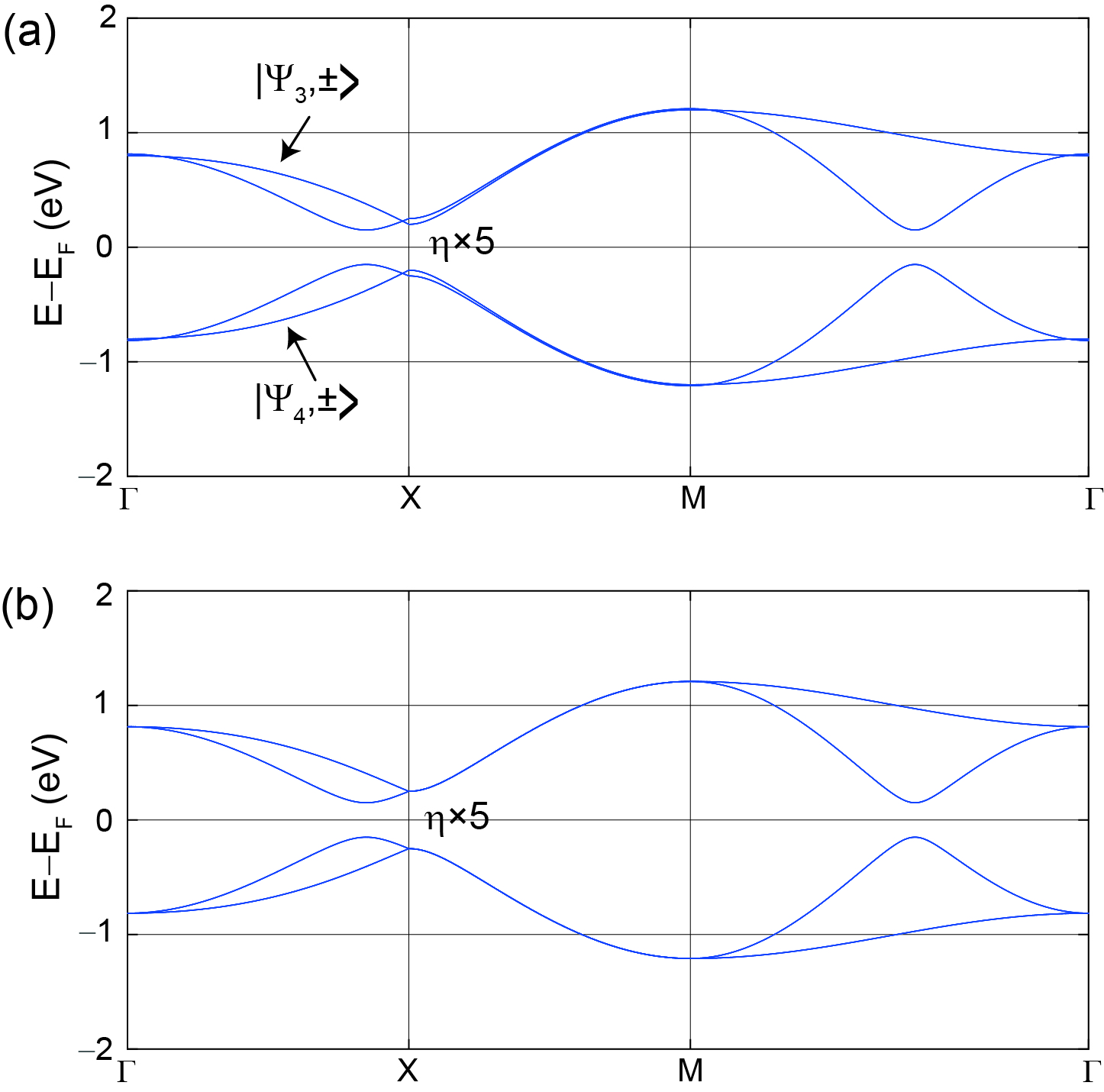}
\caption{\label{fig:f_8n}(a) Electronic dispersion of the $p_xp_y$ model using the parameters of Ref.~\citep{klemenz2020systematic}, with $\mathcal{P}^{low}$ added to its low-energy block. The eigenvectors associated to the high energy bands are also indicated. (b) Electronic dispersion of $\mathcal{H}_{p_xp_y+SOC}$. Both, low ang high energy bands have been effectively perturbed by Zr's on-site SOC. $\eta$ was multiplied by a factor of five to make the splitting in subplot (a) along the $X-M$ line more visible.}
\end{figure}

\begin{figure}[tb]
\centering
\includegraphics[width=0.48\textwidth]{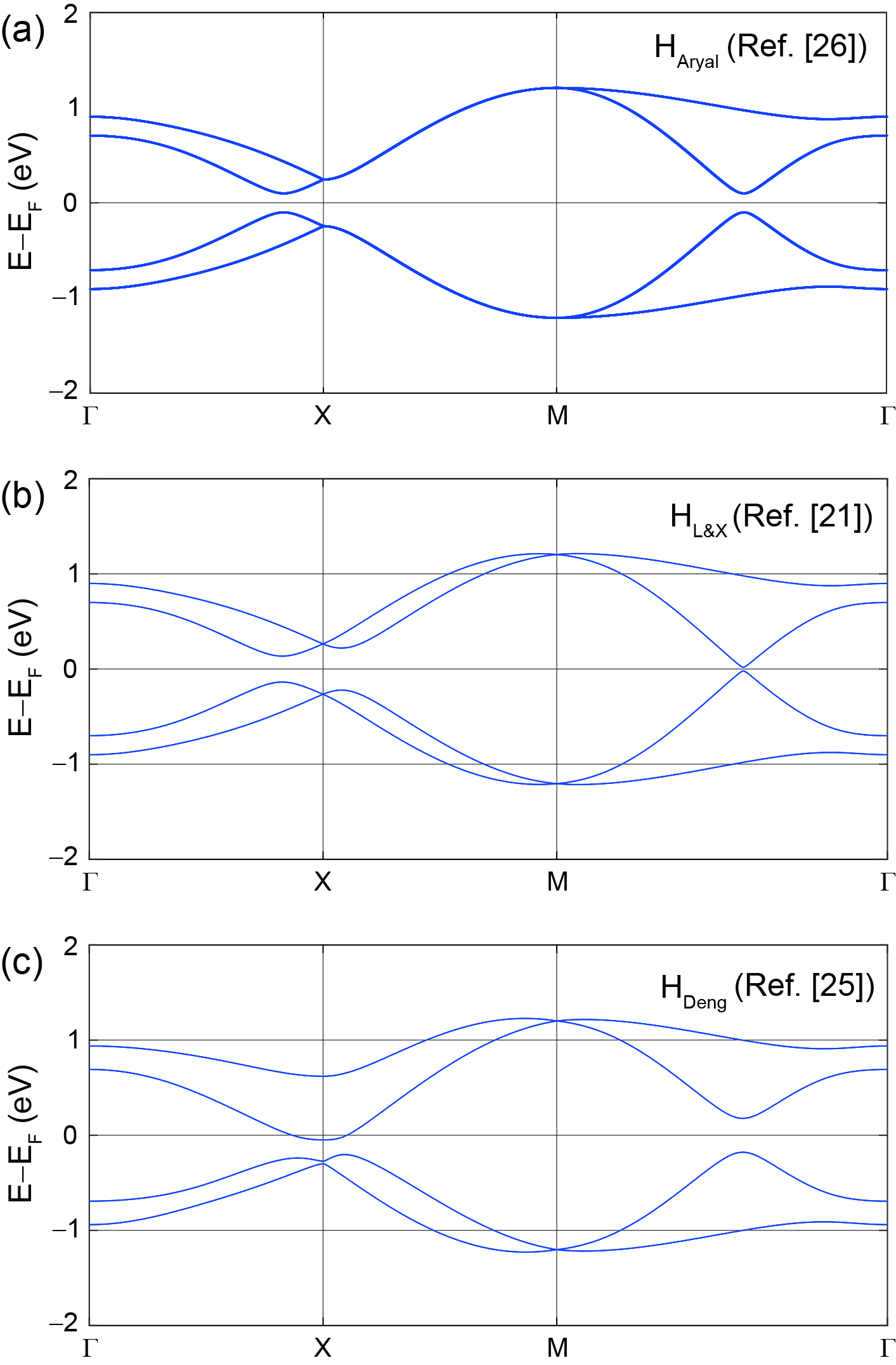}
\caption{\label{fig:pxpySOCbands}  Band structure of $p_xp_y$ models with SOC and asymmetric sublattice perturbations. (a) On-site asymmetry (Eq.~\ref{eq:Aryal}), (b) Buckling (Eq.~\ref{eq:LX}), and in-plane displacement $\delta a$ (Eq.~\ref{eq:Deng}). The TB parameters of Ref.~\cite{klemenz2020systematic} were used, along with $\lambda_p=0.4$ eV, $\epsilon=0.1$ eV and $\delta=0.1$. Our result, informed by DFT and shown in Fig.~\ref{fig:f_8n}(b), is different to all previous models.}
\end{figure}

To be specific, let $U$ and $V
$ be the unitary matrices that diagonalize the unperturbed $p_xp_y$ and \textit{16-orbital} Hamiltonians, respectively, such that the eigenvalues are ordered at $E_F$, with eigenvectors $\ket{\psi_1,\pm}$ and $\ket{\psi_2,\pm}$ (plus and minus signs label the two orthogonal spin components) (Fig.~\ref{fig:f_7n}(a)) taking the uppermost four entries:

\begin{align}
\mathcal{H}_{p_xp_y} &= U^\dagger H_{p_xp_y}\otimes\sigma_0 U,\text{ and}\nonumber\\
\mathcal{H}_{16} &= V^\dagger H_{16}\otimes\sigma_0 V.
\end{align}

To add SOC to $\mathcal{H}_{16}$ we must express it in the same basis:
\begin{align}
\mathcal{P} &= V^\dagger H^{Zr}_{SOC} V,
\end{align}
where the superscript in $H^{Zr}_{SOC}$ emphasizes that it is only Zr on-site SOC what is being taken into account. As the low-energy bands are gapped under Zr on-site SOC in the 16-orbital model, its low-energy block will be perturbed, however, $\mathcal{P}$ in general will not be block diagonal, which implies that $\mathcal{H} = \mathcal{H}_{16} + \mathcal{P}$ will not be block diagonal either.

L\"{o}wdin partitioning method transforms arbitrary perturbations into a block-diagonal form, nevertheless, so that the vector spaces associated to the low and high energy bands are uncoupled again. Furthermore, the zeroth-order L\"{o}wdin partition scheme is to neglect non-block-diagonal terms, giving rise to the gapped electronic dispersion of the low-energy block shown in Fig.~\ref{fig:f_7n}(b), which is a promising step towards a 4-orbital $p_xp_y$ model with SOC. Our numerical results lead to a SOC perturbation of the low-energy matrix block of the 16-orbital Hamiltonian with SOC [41]:
\begin{align}
\mathcal{P}^{low} &=
\begin{pmatrix}
0 & 0 & 0 & \eta \\
0 & 0 & \eta & 0 \\
0 & \eta^* & 0 & 0 \\
\eta^* & 0 & 0 & 0
\end{pmatrix}
,
\end{align}
at the vicinity of the band crossing, and in the basis of eigenvectors $\{\ket{\psi_1,+}, \ket{\psi_1,-}, \ket{\psi_2,+}, \ket{\psi_2,-}\}$, which are explicitly provided in Table \ref{tab:psi12}. Here, $\eta$ is a complex parameter that depends on $\mathbf{k}$; $\eta(\mathbf{k}_{16}^{cross})=(2+30i)$ meV. We add this perturbation onto the low-energy block $\mathcal{H}_{p_xp_y}^{low}$ of the $p_xp_y$ model in Fig.~\ref{fig:f_7n}(c), {\em thus finally embedding it with a DFT-based, effective SOC coupling}.

\begin{table}
\centering
\caption{Orbital character of eigenvectors $\ket{\psi_1}$ and $\ket{\psi_2}$ for 16-orbital model and $p_xp_y$ model in the vicinity of the band crossing. The $\ket{d}$ orbitals belong to Zr, and the rest to Si.}
\begin{tabular}{l|c|c|c|c}
\hline
\hline
Orbital & $|\psi_1^{(16)}\rangle$ & $|\psi_2^{(16)}\rangle$ & $|\psi_1^{(p_xp_y)}\rangle$ & $|\psi_2^{(p_xp_y)}\rangle$\\
\hline
$|d_{xy}^B\rangle$      & 0.000  & 0.021-0.284i & -            & - \\
$|d_{yz}^B\rangle$      & 0.000  &-0.561-0.041i & -            & - \\
$|d_{xz}^B\rangle$      &-0.063i & 0.000        & -            & - \\
$|d_{x^2-y^2}^B\rangle$ & 0.569  & 0.000        & -            & - \\
$|d_{xy}^A\rangle$      & 0.000  & 0.021-0.284i & -            & - \\
$|d_{yz}^A\rangle$      & 0.000  & 0.561+0.041i & -            & - \\
$|d_{xz}^A\rangle$      &-0.063i & 0.000        & -            & - \\
$|d_{x^2-y^2}^A\rangle$ &-0.569  & 0.000        & -            & - \\
$|s^A\rangle$           &-0.182  & 0.000        & -            & - \\
$|p_x^A\rangle$         & 0.337i & 0.000        & -0.787+0.618i& 0.000 \\
$|p_y^A\rangle$         & 0.000  & 0.318+0.023i & 0.000        & 0.787-0.618i\\
$|p_z^A\rangle$         &-0.160  & 0.000        & -            & - \\
$|s^B\rangle$           & 0.182  & 0.000        & -            & - \\
$|p_x^B\rangle$         &-0.337i & 0.000        & 0.787+0.618i & 0.000 \\
$|p_y^B\rangle$         & 0.000  & 0.318+0.023i & 0.000        & 0.787+0.618i \\
$|p_z^B\rangle$         &-0.160  & 0.000        & -            & - \\
\hline
\hline
\end{tabular}
\label{tab:psi12}
\end{table}

Now, given that the SOC perturbation only affected the bands that produced the crossing, the remaining two bands in the $p_xp_y$ dispersion are still unmodified. In Fig.~\ref{fig:f_8n}(a), we show the bands of the $p_xp_y$ model once $\mathcal{P}^{low}$ has been added to its low-energy block. L\"{o}wdin procedure only addresses how the dispersion around $E_F$ is modified, however, a characteristic of ZrSiS is the symmetry-protected crossing at $X$, which is not observed in Fig.~\ref{fig:f_8n}(a). To recover such degeneracy, the two high-energy bands in the $p_xp_y$ model (Fig.~\ref{fig:f2}) must also be modified. Analogously to the low-energy block, we add an high-energy SOC Hamiltonian:
\begin{align}
\mathcal{P}^{high} &=\mathcal{P}^{low}
\end{align}
which is now written in the basis $\{\ket{\psi_3,+}$, $\ket{\psi_3,-}$, $\ket{\psi_4,+}$, $\ket{\psi_4,-}\}$. Thus, the effective on-site SOC inherited from Zr in the $p_xp_y$ model can be expressed as:
\begin{align}
\mathcal{P}_{p_xp_y} =
\begin{pmatrix}
\mathcal{P}^{low} & O \\
O & \mathcal{P}^{low}
\end{pmatrix},
\end{align}
and the Hamiltonian under the effective on-site SOC expressed in the eigenbasis of the unperturbed $p_xp_y$ Hamiltonian happens to be:
\begin{align}
\mathcal{H}_{p_xp_y+SOC} = \mathcal{H}_{p_xp_y} + \mathcal{P}_{p_xp_y},
\end{align}
with an associated dispersion degenerate at $X$ depicted in Fig~\ref{fig:f_8n}(b) (see a similar degeneracy in Figs.~\ref{fig:f_3n}(c),~\ref{fig:f_3n}(d),~\ref{fig:f_5n}(b), and~\ref{fig:f_5n}(d)).

The last question is whether SOC makes the Si $A$ and $B$ sublattices inequivalent. To answer this, we transformed the $p_xp_y$ SOC Hamiltonian $\mathcal{P}_{p_xp_y}$  back into the original basis ($\{\ket{p_x^A,+},\ket{p_y^A,+},\ket{p_x^B,+},\ket{p_y^B,+}$, $\ket{p_x^A,-},\ket{p_y^A,-},\ket{p_x^B,-},\ket{p_y^B,-}\}$), to get:
\begin{align}
P_{p_xp_y} &= U\mathcal{P}_{p_xp_y}U^\dagger
\nonumber\\
&=
\begin{pmatrix}
O & Q \\
Q^\dagger & O
\end{pmatrix},
\end{align}
where $Q$ has the form:
\begin{align}
\label{eq:Q}
Q =
\left(
\begin{smallmatrix}
0 & -\operatorname{Re}(\eta) & 0 & -i\operatorname{Im}(\eta) \\
-\operatorname{Re}(\eta) & 0 & -i\operatorname{Im}(\eta) & 0 \\
0 & i\operatorname{Im}(\eta) & 0 & \operatorname{Re}(\eta) \\
i\operatorname{Im}(\eta) & 0 & \operatorname{Re}(\eta) & 0
\end{smallmatrix}
\right)
\end{align}
along the $\Gamma-X$ segment: {\em Zr's on-site SOC did produce an inequivalence among A and B sublattices on the $p_xp_y$ model}, as the hopping from $\ket{p_x^A,\pm}$ to $\ket{p_y^A,\pm}$ has opposite sign with respect to the hopping from $\ket{p_x^B,\pm}$ to $\ket{p_y^B,\pm}$ at the crossing at $\Gamma-X$. Thus, Zr is responsible for the lowering of symmetry that Refs.~\cite{luo2015room} and \cite{deng2022twisted} assigned to the atoms within their square nets, and which Ref.~\cite{aryal2022topological} postulated {\em ad hoc}. { Fig.~\ref{fig:pxpySOCbands} displays the band structures of the $p_xp_y$ models with SOC plus the sublattice asymmetric perturbations discussed in Sec.~\ref{sec:KSC} for comparison with our model.

The correct, actual SOC Hamiltonian for ZrSiS--obtained with DFT guidance here--can also be expressed in terms of the Pauli matrices introduced in Sec.~\ref{sec:KSC}, and it is our main contribution:
\begin{align}
P_{p_xp_y} = \operatorname{Re}(\eta)\tau_x\otimes\upsilon_z\otimes\sigma_x + \operatorname{Im}(\eta)\tau_x\otimes\upsilon_y\otimes\sigma_x.
\end{align}

The picture that arises as a result is as follows: Fig~\ref{fig:TS} displays the trajectories that an electron originally with spin up follows to connect with atoms on the same sublattice. The electron will first hop to a Zr atom, which will be in charge of switching its spin through on-site SOC. Then, it will hope again to a Si atom of the same sublattice to produce an effective hopping term that acquires the value of $-\operatorname{Re}(\eta)$ in the vicinity of $\mathbf{k}_{cross}$. Observe that, {\em depending on the direction of hopping} (horizontal {\em versus} vertical), the electron will pass either {\em under} or {\em over} the Si layer. Since the relative positions of neighboring Zr atoms depends on the sublattice, the presence of Zr facilitates the asymmetry observed in Eq.~\eqref{eq:Q}.
\begin{figure}[t]
\centering
\includegraphics[width=0.48\textwidth]{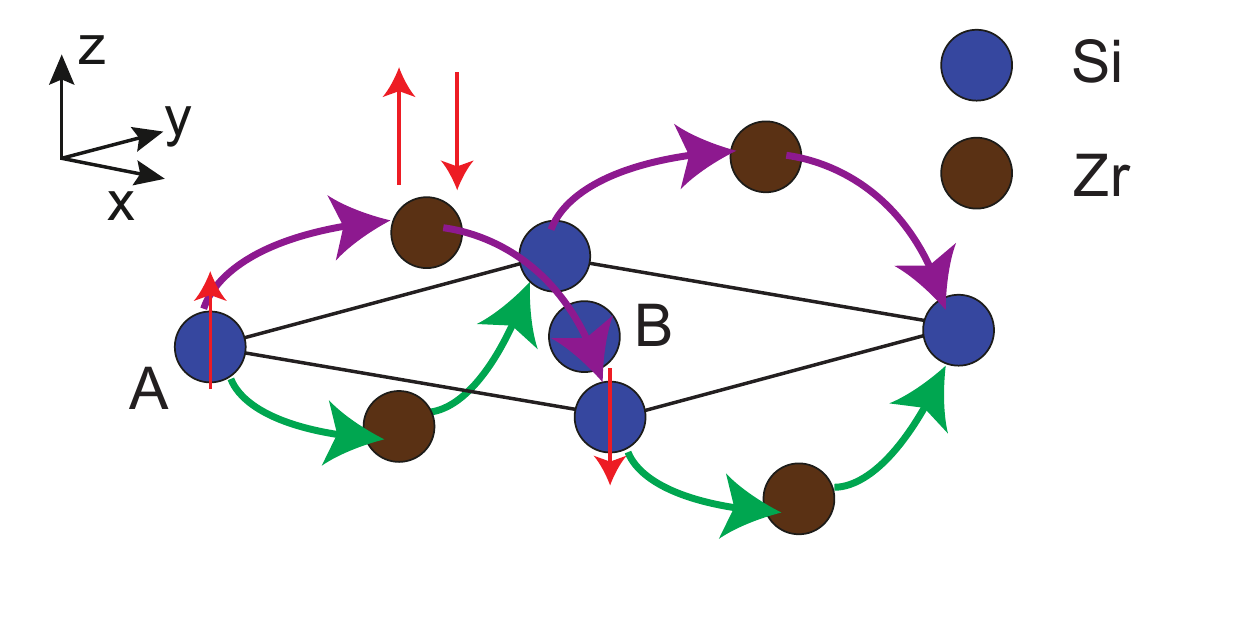}
\caption{\label{fig:TS}Connection between $\ket{p_x}$ and $\ket{p_y}$ orbitals with opposite spins through Zr's on-site SOC.}
\end{figure}

Informed by DFT, we have systematically showed how to embed Zr on-site SOC into the $p_xp_y$ model for ZrSiS. In this manner, we not only connect the observations made in previous works, but also, we show that on-site SOC does not require the addition of extra terms to gap the dispersion around the Fermi level.}

\section{\label{sec:Con}Conclusions}

In this work, we followed a systematic procedure to induce a gap in the $p_xp_y$ model. Relying on DFT all along, we concluded that (i) on-site SOC is the main responsible to the SOC-gaps reported on the band structure of ZrSiS around $E_F$. Furthermore, (ii) we determined that Zr's orbitals play a key role on the on-site SOC, and that coupling among opposite spins is crucial to split degenerate bands around $E_F$. Then, (iii) we developed a DFT-based TB model larger than $p_xp_y$, but smaller than the DZP one, which allowed to treat SOC carefully. At that point, (iv) we applied a L\"odwin decomposition to extract an effective SOC, and ended up with a $p_xp_y$ dispersion gapped at $E_F$ and consistent with DFT, without arbitrary lowering the symmetry of the square net, nor postulating the entries of the SOC interaction in an {\em ad hoc} manner. This model extends the usefulness of the minimal square-net nodal line semimetal $p_xp_y$ TB models.

\acknowledgments

G.S.O.G.~and S.B.L.~acknowledge partial funding from the MonArk NSF Quantum Foundry, supported by the National Science Foundation Q-AMASE-i program under NSF award No. DMR-1906383. Calculations were performed at the University of Arkansas' Pinnacle supercomputer, funded by the U.S. National Science Foundation, the Arkansas Economic Development Commission, and the Office of the Vice Provost for Research and Innovation. A.G.F. was funded by project PID2022-137078NB-I00 (MCIU/AEI/FEDER, EU) and Asturias FICYT under Grant No. AYUD/2021/51185 with the support of FEDER funds.
 We thank A.~Huam\'an for his assistance with the L\"odwin projection scheme. Conversations with A.~Fereidouni Ghaleh Minab, J.~Hu, and J.~Ferrer are gratefully acknowledged.


\providecommand{\noopsort}[1]{}\providecommand{\singleletter}[1]{#1}%

\appendix

\section{Tight binding description from DFT}\label{appendixA}

To construct a TB model, it is necessary to compute two-center integrals of the form \cite{slater1954simplified}:
\begin{align}
S_{\mu\nu} &= \int\psi_\mu(\mathbf{r}-\mathbf{R}_{\mu})^*\psi_\nu(\mathbf{r}-\mathbf{R}_{\nu}),
\\
H_{\mu\nu} &= \int\psi_\mu(\mathbf{r}-\mathbf{R}_{\mu})^*\hat{H}\psi_\nu(\mathbf{r}-\mathbf{R}_{\nu}),
\end{align}
where $\mathbf{R}_\nu$ and $\mathbf{R}_\mu$ are atomic positions in which localized orbitals $\psi_\mu$ and $\psi_\nu$ are centered at, respectively.

SIESTA \cite{soler2002siesta} constructs an auxiliary supercell, and then proceeds to calculate the two center integrals between each orbital within the u.c.~with every orbital in the supercell. The information of these integrals, written into text files, is used to explicitly build the entries of the Hamiltonian for each $k-$point in reciprocal space here:
\begin{equation}
H_{\mu\nu}(\mathbf{k}) = \sum_{\mu'\equiv\mu}H_{\mu'\nu}e^{i(\mathbf{R}_\mu-\mathbf{R}_\nu)\cdot\mathbf{k}},
\end{equation}
where the primed index runs over the full supercell and $\mu'\equiv\mu$ implies that $\psi_\mu$ is the unit cell counterpart of $\psi_{\mu'}$.

For elements containing up to $d-$electrons, Slater and Koster proposed that the values of these two center integrals can be expressed as \cite{slater1954simplified,martin_2020}:
\begin{align}
H_{\mu\nu} = t_\sigma f_\sigma(l,m,n) + t_\pi f_\pi(l,m,n) + t_\delta f_\delta(l,m,n)
\label{eq:SlaterK}
\end{align}
where $t_\alpha$ are the known TB parameters while $f_\alpha$ are functions of the projection cosines $l$, $m$ and $n$ that can be explicitly found in the original work of Slater and Koster \cite{slater1954simplified}.

Once positions of the atoms and the values of the two-center integrals $H_{\mu'\nu}$ are known, the TB parameters associated to the orbitals $\psi_\nu$ and $\psi_\mu$ can be obtained by solving the system of linear equations of the form of Eqn.~\eqref{eq:SlaterK}, generated by $\psi_\nu$ and the set of orbitals in the supercell whose equivalent orbital in the u.c.~is $\psi_\mu$. Typically, these systems will be overdetermined--there are more neighbors than unknown parameters. There can be instances where $f_\sigma(l,m,n) = f_\pi(l,m,n) = f_\delta(l,m,n) = 0$ and still $H_{\mu'\nu} \neq 0$. Even though the contribution of such integrals cannot be considered in the Slater-Koster approximation, excluding them does not brake any symmetries and we have found that the change in the dispersion of the bands is negligible.

In the following subsection, we report a TB description of the 16-orbital Hamiltonian. The parameters used to build the model were obtained by solving systems of linear equations as described in the present subsection.

\subsection{16-orbital model}

To produce the dispersion depicted in Fig.~\ref{fig:f_5n}(d), we label the atoms in the u.c.~first. In this model, S orbitals have already been excluded from the basis set. Fig.~\ref{fig:AC} illustrates our labelling convention.
We define the Hamiltonian matrix that takes into account interactions in sub-blocks of $4\times4$ matrices:
\begin{align}
H_{16} =
\begin{pmatrix}
H^{(2)} & O & H^{(7)} & H^{(6)}\\
O & H^{(2)} & H^{(5)} & H^{(4)}\\
H^{(7)\dagger} & H^{(5)\dagger} & H^{(1)} & H^{(3)}\\
H^{(6)\dagger} & H^{(4)\dagger} & H^{(3)\dagger} & H^{(1)}
\end{pmatrix}.
\end{align}
Each block contains the information of the interaction of all the orbitals of one atom with every orbital of another atom. The order of the block rows and columns is ZrB, ZrA, SiA, SiB (i.e. $H^{(5)}$ contains the information of the interaction among ZrA and SiA orbitals).

Using the Slater-Koster approach, we wrote the entries of each block in terms of the lattice vectors (Table \ref{LatticeV}), the displacement vectors $\mathbf{r}_j$ (Table \ref{DispV}), and the TB parameters $t_j$ (Table \ref{TBparameters}).
Recalling that the projection cosines $l$, $m$, $n$, along $x$, $y$, $z$ respectively are defined by: $$l_j = \frac{r_{jx}}{|\mathbf{r}_j|},\quad m_j = \frac{r_{jy}}{|\mathbf{r}_j|},\quad n_j = \frac{r_{jz}}{|\mathbf{r}_j|},$$the non-zero matrix elements of the Hamiltonian expressed in the basis:
$\{$Si: $s$, $p_x$, $p_y$, $p_z$; Zr: $d_{xy}$, $d_{yz}$, $d_{xz}$, $d_{x^2-y^2}\}$, are written in Table\ref{tab:Ham16}.

\begin{table}
\centering
\caption{Lattice constants and lattice vectors for ZrSiS.}
\begin{tabular}{c}
\hline
\hline
Lattice constants\\
\hline
$a = 3.554\,\AA$, $c = 8.055\,\AA$\\
\hline
Lattice vectors\\
\hline
$\mathbf{a}_1 = a(1,0,0)$, $\mathbf{a}_2 = a(0,1,0)$,  $\mathbf{a}_3 = c(0,0,1)$\\
\hline
\hline
\end{tabular}
\label{LatticeV}
\end{table}

\begin{table}
\centering
\caption{Displacement vectors that connect the indicated atomic species.}
\begin{tabular}{ c }
 \hline
 \hline
SiA-SiB \\
\hline
$\mathbf{r}_1 = \frac{1}{2}\left(\mathbf{a}_1 + \mathbf{a}_2\right)$,  $\mathbf{r}_2 = \frac{1}{2}\left(-\mathbf{a}_1 + \mathbf{a}_2\right)$,\\
$\mathbf{r}_3 = -\frac{1}{2}\left(\mathbf{a}_1 + \mathbf{a}_2\right)$,  $\mathbf{r}_4 = \frac{1}{2}\left(\mathbf{a}_1 - \mathbf{a}_2\right)$ \\
\hline
ZrA-SiA \\
\hline
$\mathbf{r}_5 = \frac{1}{2}\mathbf{a}_1 - \frac{1}{4}\mathbf{a}_3$,  $\mathbf{r}_6 = -\frac{1}{2}\mathbf{a}_1 - \frac{1}{4}\mathbf{a}_3$,\\
\hline
ZrA-SiB \\
\hline
$\mathbf{r}_7 = \frac{1}{2}\mathbf{a}_2 - \frac{1}{4}\mathbf{a}_3$,  $\mathbf{r}_8 = -\frac{1}{2}\mathbf{a}_2 - \frac{1}{4}\mathbf{a}_3$,\\
\hline
ZrB-SiA \\
\hline
$\mathbf{r}_9 = \frac{1}{2}\mathbf{a}_2 + \frac{1}{4}\mathbf{a}_3$,  $\mathbf{r}_{10} = -\frac{1}{2}\mathbf{a}_2 + \frac{1}{4}\mathbf{a}_3$,\\
\hline
ZrB-SiB \\
\hline
$\mathbf{r}_{11} = \frac{1}{2}\mathbf{a}_1 + \frac{1}{4}\mathbf{a}_3$,  $\mathbf{r}_{12} = -\frac{1}{2}\mathbf{a}_1 + \frac{1}{4}\mathbf{a}_3$,\\
\hline
Zr-Zr and Si-Si \\
\hline
$\mathbf{r}_{13} = \mathbf{a}_1$, $\mathbf{r}_{14} = \mathbf{a}_2$, $\mathbf{r}_{15} = -\mathbf{a}_1$, $\mathbf{r}_{16} = -\mathbf{a}_2$ \\
\hline
\hline
\end{tabular}
\label{DispV}
\end{table}

\begin{table}
\centering
\caption{On-site energies $E_j$ and tight binding parameters $t_j$ for the 16-orbital model. We have used $\alpha$ and $\beta$ ($\alpha\neq\beta$) to represent both $x$ and $y$ subscripts. Additionally, $\sigma$ is used if the lobes of the orbital are oriented in the direction of the displacement vector that links the two atomic sites, while $\pi$ is used when the lobes are perpendicular to it.}.
\begin{tabular}{ c|c|c|c|c|c|c }
\hline
\hline
$j$ & \multicolumn{2}{c|}{Atom} & $r$ ($\AA$) & \multicolumn{2}{c|}{Orbital} & $E_j$ ($eV$)\\
 \hline
1 & \multicolumn{2}{c|}{\multirow{3}{1.3em}{Zr}} &	\multirow{3}{2.8em}{0.0000}	&	\multicolumn{2}{c|}{$d_{xy}$} &	3.199	\\
2 & \multicolumn{2}{c|}{} &		&	\multicolumn{2}{c|}{$d_{\alpha z}$} &	3.220	\\
3 & \multicolumn{2}{c|}{} &		&	\multicolumn{2}{c|}{$d_{x^2-y^2}$} &	3.239	\\
\hline
4 & \multicolumn{2}{c|}{\multirow{3}{1.3em}{Si}}	&	\multirow{3}{2.8em}{0.0000}	&	\multicolumn{2}{c|}{$s$}	&	-5.179	\\
5 & \multicolumn{2}{c|}{} &		&	\multicolumn{2}{c|}{$p_\alpha$}	&	2.301	\\
6 & \multicolumn{2}{c|}{} &		&	\multicolumn{2}{c|}{$p_z$}	&	2.564	\\
\hline
$j$ & AtomR & AtomC & $r$ ($\AA$) & OrbR & OrbC & $t_j$ ($eV$) \\
\hline
1 & \multirow{6}{1.3em}{Si}	&	\multirow{6}{1.3em}{Si}	&	\multirow{6}{2.8em}{2.5062}	&	$s$	&	$s$	&	-1.690	\\
2 & 	&		&		&	$s$	&	$p_\alpha$	&	-2.207	\\
3 & 	&		&		&	$p_\alpha$	&	$s$	&	2.207	\\
4 & 	&		&		&	$p_\alpha$	&	$p_\alpha$	&	1.691	\\
5 & 	&		&		&	$p_\alpha$	&	$p_\beta$	&	3.046	\\
6 & 	&		&		&	$p_z$	&	$p_z$	&	-0.680	\\
\hline
7 & \multirow{8}{1.3em}{Zr}	&	\multirow{8}{1.3em}{Si}	&	\multirow{8}{2.8em}{2.8210}	&	$d_{xy}$	&	$p_\pi$	&	0.522	\\
8 & 	&		&		&	$d_{\sigma z}$	&	$s$	&	1.141	\\
9 & 	&		&		&	$d_{\sigma z}$	&	$p_\sigma$	&	1.425	\\
10 & 	&		&		&	$d_{\sigma z}$	&	$p_z$	&	-1.685	\\
11 & 	&		&		&	$d_{\pi z}$	&	$p_\pi$	&	-0.535	\\
12 & 	&		&		&	$d_{x^2-y^2}$	&	$s$	&	-1.128	\\
13 & 	&		&		&	$d_{x^2-y^2}$	&	$p_\sigma$	&	-0.634	\\
14 & 	&		&		&	$d_{x^2-y^2}$	&	$p_z$	&	2.168	\\
\hline
15 & \multirow{3}{1.3em}{Si}	&	\multirow{3}{1.3em}{Si}	&	\multirow{3}{2.8em}{3.5440}	&	$s$	&	$s$	&	-0.139 \\
16 & 	&		&		&	$s$	&	$p_\alpha$	&	-0.279	\\
17 & 	&		&		&	$p_\alpha$	&	$p_\alpha$	&	0.555 \\
\hline
18 & Zr	&	Zr	&	3.6460	&	$d_{x^2-y^2}$	&	$d_{x^2-y^2}$	&	-0.407	\\
\hline
\hline
\end{tabular}
\label{TBparameters}
\end{table}

\begin{table}
\centering
\caption{16-orbital model Hamiltonian entries.}
\begin{tabular}{ c }
\hline
\hline
SiA-SiB ($j=1,2,3,4$)\\
\hline
$H_{1,4}^{(3)} = t_1\sum_{j} e^{i\mathbf{k}\cdot\mathbf{r_j}}$,
\\
$H_{1,2}^{(3)} = t_2\sum_{j} l_je^{i\mathbf{k}\cdot\mathbf{r_j}}$,
\\
$H_{1,3}^{(3)} = t_2\sum_{j} m_je^{i\mathbf{k}\cdot\mathbf{r_j}}$,
\\
$H_{2,1}^{(3)} = t_3\sum_{j} l_je^{i\mathbf{k}\cdot\mathbf{r_j}}$,
\\
$H_{2,2}^{(3)} = t_4\sum_{j} l_j^2e^{i\mathbf{k}\cdot\mathbf{r_j}}$,
\\
$H_{2,3}^{(3)} = H_{3,2}^{(5)} = t_5\sum_{j} l_jm_je^{i\mathbf{k}\cdot\mathbf{r_j}}$,
\\
$H_{3,1}^{(3)} = t_3\sum_{j} m_je^{i\mathbf{k}\cdot\mathbf{r_j}}$,
\\
$H_{3,3}^{(3)} = t_4\sum_{j} m_j^2e^{i\mathbf{k}\cdot\mathbf{r_j}}$,
\\
$H_{4,4}^{(3)} = t_6\sum_{j} (1 - n_j^2)e^{i\mathbf{k}\cdot\mathbf{r_j}}$
\\
\hline
ZrA-SiA ($j=5,6$;$h=5$), ZrB-SiB ($j=11,12$;$h=6$)\\
\hline
$H_{1,3}^{(h)} = t_7\sum_{j} l_j(1 - 2m_j^2)e^{i\mathbf{k}\cdot\mathbf{r_j}}$,
\\
$H_{2,3}^{(h)} = t_{11}\sum_{j} n_j(1 - 2m_j^2)e^{i\mathbf{k}\cdot\mathbf{r_j}}$,
\\
$H_{3,1}^{(h)} = t_8\sum_{j} \sqrt{3} l_jn_je^{i\mathbf{k}\cdot\mathbf{r_j}}$,
\\
$H_{3,2}^{(h)} = t_9\sum_{j} \sqrt{3} l_j^2n_je^{i\mathbf{k}\cdot\mathbf{r_j}}$,
\\
$H_{3,4}^{(h)} = t_{10}\sum_{j} \sqrt{3} n_j^2l_je^{i\mathbf{k}\cdot\mathbf{r_j}}$,
\\
$H_{4,1}^{(h)} = t_{12}\sum_{j} \frac{\sqrt{3}}{2}(l_j^2 - m_j^2) e^{i\mathbf{k}\cdot\mathbf{r_j}}$,
\\
$H_{4,2}^{(h)} = t_{13}\sum_{j} \frac{\sqrt{3}}{2}l_j(l_j^2 - m_j^2) e^{i\mathbf{k}\cdot\mathbf{r_j}}$,
\\
$H_{4,4}^{(h)} = t_{14}\sum_{j} \frac{\sqrt{3}}{2}n_j(l_j^2 - m_j^2) e^{i\mathbf{k}\cdot\mathbf{r_j}}$
\\
\hline
ZrA-SiB ($j=7,8$;$h=4$), ZrB-SiA ($j=9,10$;$h=7$)\\
\hline
$H_{1,2}^{(h)} = t_7\sum_{j} m_j(1 - 2l_j^2)e^{i\mathbf{k}\cdot\mathbf{r_j}}$,
\\
$H_{2,1}^{(h)} = t_8\sum_{j} \sqrt{3} m_jn_je^{i\mathbf{k}\cdot\mathbf{r_j}}$,
\\
$H_{2,3}^{(h)} = t_9\sum_{j} \sqrt{3} m_j^2n_je^{i\mathbf{k}\cdot\mathbf{r_j}}$,
\\
$H_{2,4}^{(h)} = t_{10}\sum_{j} \sqrt{3} m_jn_j^2e^{i\mathbf{k}\cdot\mathbf{r_j}}$,
\\
$H_{3,2}^{(h)} = t_{11}\sum_{j} n_j(1 - 2l_j^2)e^{i\mathbf{k}\cdot\mathbf{r_j}}$,
\\
$H_{4,1}^{(h)} = t_{12}\sum_{j} \frac{\sqrt{3}}{2}(l_j^2 - m_j^2) e^{i\mathbf{k}\cdot\mathbf{r_j}}$,
\\
$H_{4,3}^{(h)} = t_{13}\sum_{j} \frac{\sqrt{3}}{2}m_j(l_j^2 - m_j^2) e^{i\mathbf{k}\cdot\mathbf{r_j}}$,
\\
$H_{4,4}^{(h)} = t_{14}\sum_{j} \frac{\sqrt{3}}{2}n_j(l_j^2 - m_j^2) e^{i\mathbf{k}\cdot\mathbf{r_j}}$,
\\
\hline
Zr-Zr and Si-Si ($j=13,14,15,16$)\\
\hline
$H_{1,1}^{(1)} = E_1 + t_{15}\sum_{j}e^{i\mathbf{k}\cdot\mathbf{r_j}}$,
\\
$H_{1,2}^{(1)} = t_{16}\sum_{j=13}^{16}l_je^{i\mathbf{k}\cdot\mathbf{r_j}}$,
\\
$H_{1,3}^{(1)} = t_{16}\sum_{j}m_je^{i\mathbf{k}\cdot\mathbf{r_j}}$,
\\
$H_{2,2}^{(1)} = E_2 + t_{17}\sum_{j}l_j^2e^{i\mathbf{k}\cdot\mathbf{r_j}}$,
\\
$H_{3,3}^{(1)} = E_2 + t_{17}\sum_{j}m_j^2e^{i\mathbf{k}\cdot\mathbf{r_j}}$,
\\
$H_{4,4}^{(1)} = E_3$,
\\
$H_{1,1}^{(2)} = E_4$,
\\
$H_{2,2}^{(2)} = H_{3,3}^{(2)} = E_5$,
\\
$H_{4,4}^{(2)} = E_6 + t_{18}\sum_{j}\frac{3}{4}(l_j^2-m_j^2)e^{i\mathbf{k}\cdot\mathbf{r_j}}$
\\
\hline
\hline
\end{tabular}
\label{tab:Ham16}
\end{table}

\end{document}